\let\oldvec\vec
\documentclass[]{aa}
\let\vec\oldvec

\pdfoutput=0
\usepackage{graphicx}
\usepackage{amssymb,amsmath}
\usepackage{txfonts}
\usepackage{natbib}
\usepackage{rotating}
\usepackage{color, colortbl}

\renewcommand{\vec}[1]{\mathbf{#1}}

\definecolor{LightCyan}{rgb}{0.88,1,1}

\definecolor{Gray}{gray}{0.9}

\begin{document}


   \title{The Low-High-Low Trend of Type III Radio Burst Starting Frequencies and Solar Flare Hard X-rays}


   \author{Hamish A. S. Reid\inst{1,2}, Nicole Vilmer\inst{1}, and Eduard P. Kontar\inst{2}}
   \institute
   {$^{1}$LESIA, Observatoire de Paris, CNRS, UPMC, Universit\'{e} Paris-Diderot, 5 place Jules Janssen, 92195 Meudon Cedex, France \\
   $^{2}$SUPA School of Physics and Astronomy,   University of Glasgow, G12 8QQ, United Kingdom}
   
   \date{}

\abstract
{}
{Using simultaneous X-ray and radio observations from solar flares, we investigate the link between the type III radio burst starting frequency and hard X-ray spectral index.  For a proportion of events the relation derived between the starting height (frequency) of type III radio bursts and the electron beam velocity spectral index (deduced from X-rays) is used to infer the spatial properties (height and size) of the electron beam acceleration region.  Both quantities can be related to the distance travelled before an electron beam becomes unstable to Langmuir waves.}
{ To obtain a list of suitable events we considered the RHESSI catalogue of X-ray flares and the Phoenix 2 catalogue of type III radio bursts.  From the 200 events that showed both type III and X-ray signatures, we selected 30 events which had simultaneous emission in both wavelengths, good signal to noise in the X-ray domain and $>$ 20 seconds duration.}
{We find that $>50~\%$ of the selected events show a good correlation between the starting frequencies of the groups of type III bursts and the hard X-ray spectral indices.  A low-high-low trend for the starting frequency of type III bursts is frequently observed.  Assuming a background electron density model and the thick target approximation for X-ray observations, this leads to a correlation between starting heights of the type III emission and the beam electron spectral index.  Using this correlation we infer the altitude and vertical extents of the flare acceleration regions.  We find heights from 183 Mm down to 25 Mm while the sizes range from 13 Mm to 2 Mm.  These values agree with previous work that places an extended flare acceleration region high in the corona.  We also analyse the assumptions that are required to obtain our estimates and explore possible extensions to our assumed model.  We discuss these results with respect to the acceleration heights and sizes derived from X-ray observations alone.}
{}


\titlerunning{X-ray Flare and Type III Radio Diagnostics}
\authorrunning{Reid et al}

   \maketitle

\section{Introduction} \label{intro}

Electromagnetic signatures during flares allow us to diagnose remotely what is occurring in the solar atmosphere.  We can detect the presence of non-thermally distributed electrons via an enhanced signal in radio and X-ray wavelengths.  Through a series of assumptions we can deduce properties of these electrons and the ambient environment that caused them to emit photons.  Unfortunately the spatial characteristics of the source regions for the electron energisation out of a thermal distribution remain largely unknown.

The enhanced X-ray emission in the tens of keV range and above is believed to come from non-thermal electrons in the low atmosphere during solar flares (see \citet{Fletcher_etal2011}, for an observational review).  When the electrons enter the high density plasma of the solar chromosphere they lose all their energy and thermalise via electron-ion Coulomb collisions.  Bremsstrahlung X-rays are emitted but only contain some $10^{-5}$ of the incident non-thermal electron energy \citep[see][as a recent review]{Holman_etal2011}.  Spacecraft detect the X-ray source both directly and through X-rays reflected from the solar surface \citep[e.g.][]{KontarJeffrey2010}.  We can use the X-ray signature to deduce the temporal, spatial, and energetic profile of the energised electrons \citep{Kontar_etal2011}.  A common soft-hard-soft trend \citep[e.g.][]{ParksWinckler1969,Benz1977} has been observed where the X-ray spectral index starts high, becomes low during the most intense part of the flare, and then finishes high.

Type III radio emission at frequencies $\leq 4$~GHz is believed to be caused by high energy electrons streaming through the corona and interplanetary space (see \citet{Nindos_etal2008} for a recent review).  The bump-in-tail instability causes the electrons to induce high levels of Langmuir waves in the background plasma \citep{GinzburgZhelezniakov1958}.  Non-linear wave-wave interactions then convert a small fraction of the energy contained in the Langmuir waves into electromagnetic emission near the local plasma frequency or at its harmonics \citep[e.g.][]{Melrose1980}.  Consequently, we detect radio emission drifting from high to low frequencies as the high energy electrons propagate through the corona and out into interplanetary space.  Ground based telescopes are used to detect the radio waves $\geq 10$ MHz that are able to penetrate the Earth's atmosphere.

Hard X-ray and type III radio observations are known to be statistically correlated in time during solar flares \citep{Kane1981,Raoult_etal1985,Hamilton_etal1990,Aschwanden_etal1995,ArznerBenz2005}.  Indeed recent statistical studies \citep{Saint-HilaireBenz2003,Benz_etal2005,Benz_etal2007,Michalek_etal2009} that include all types of coherent radio emission leave little doubt as to the connection between coherent radio and hard X-ray emission.  There has also been copious studies which deal with individual events \citep[e.g.][]{KaneRaoult1981,Kane_etal1982,Benz_etal1983,Dennis_etal1984,Vilmer_etal2002,Christe_etal2008} in more intricate detail.  Such studies are helpful to delve deeper into this complicated phenomenon.

\begin{figure}
 \includegraphics[width=0.99\columnwidth]{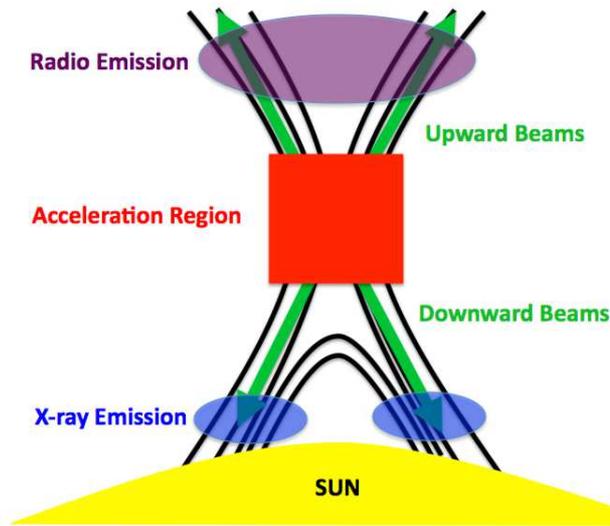}
\caption{Cartoon showing the relation between the presumed electron beam acceleration site in the corona with respect to the radio and HXR emitting sites.}
\label{fig:overview}
\end{figure}

One such study of locations of the HXR and radio sources observed for the 20 February 2002 flare \citep{Vilmer_etal2002} shows a close correspondence between the change of the HXR configuration in the 25-40 keV range and the radio source at the
highest frequency imaged with the Nan\c{c}ay Radioheliograph (410 MHz - 73 cm).  This event strongly supports the idea of a common acceleration (injection) site for HXR and radio emitting electrons located in the current sheet formed above the loop, in close agreement with the simple cartoon derived previously by \citet{Aschwanden_etal1995,AschwandenBenz1997}. Figure \ref{fig:overview} shows the simple cartoon illustrating the location of the electron beam acceleration site in the corona with respect to the locations of the radio and HXR emitting sites. It is well known however that such a simple link between HXR and radio type III sources is not always as simple as what could be derived from this simple scenario \citep[see e.g.][for reviews]{PickVilmer2008,Vilmer2012}.



Regarding the acceleration region of energetic particles in solar flares, it has been postulated recently that the coronal X-ray sources could indicate their location \citep{Krucker_etal2010, Ishikawa_etal2011, ChenPetrosian2012}.  The sources are found quite low in altitude \citep[see also][]{Xu_etal2008,Kontar_etal2011b,Guo_etal2012}, some 20~Mm above the soft X-ray loops.  The approximated volume for these sources is around $10^{26}~\rm{to}~ 10^{27}~\rm{cm}^3$ giving rough 1D estimates of $10^9$ cm (10 Mm).  It is common for coronal X-ray sources to have a higher spectral index (softer spectrum) than X-ray footpoints \citep{Emslie_etal2003, BattagliaBenz2007,KruckerLin2008}.  With current X-ray instrumentation it can be hard to observe faint sources in the presence of very strong footpoint sources and so the high densities of coronal sources could be a selection effect.

On the other hand, in the case of the simple scenario shown in Figure \ref{fig:overview}, it has been demonstrated in \citet{Reid_etal2011} how an anti-correlation between type III starting frequencies and HXR spectral index can be used to deduce the acceleration height and size of the electron beam.  Indeed, the observed anti-correlation between the starting frequency of type III bursts and the hard X-ray spectral index is used together with numerical / analytical models of electron transport in the corona to derive these quantities. Such an anti-correlation between type III starting frequencies and HXR spectral indices has been also previously reported in the literature, since it was noticed that the X-ray / type III correlation increases systematically with the peak spectral hardness of HXR emission and the type III burst starting frequency \citep{Kane1981,Hamilton_etal1990}. A simple physical explanation for this observed property is linked to the fact that through propagation effects an electron beam with a hard spectrum will excite Langmuir waves faster and closer to the beam injection site than a beam with a softer spectrum.  This faster instability onset is the explanation for the correlation between the HXR spectrum and the type III starting frequency, in the case when the two electron populations have a common origin and similar spectral properties.

In this work, we re-examine on a statistical basis the link between HXR spectra and type III starting frequencies. The aim is twofold:
\begin{itemize}
\item To estimate the proportion of events for which a simple link (as the one inferred from figure 1) is found
\item To deduce  for these events  for which an anti-correlation is found  between type III frequencies and HXR spectral indices, the characteristics of electron acceleration sizes and heights.
\end{itemize}



We start by defining our selection criteria for events in Section \ref{selection} and analysing the two observables: starting frequency of radio type III bursts and HXR spectral indices.  In Section \ref{electron_beam} we derive from these observables to starting heights of the radio emission and the electron beam spectral indices and investigate how they are related.  In Section \ref{properties} we recall the model of electron beam propagation which is used to derive from the previous quantities the values of the acceleration height and size of the energetic electron beam for each of the studied events.  We discus in Section \ref{discussion} the results as well as the assumptions of the model and discuss the flare morphology of the different events as revealed from combined X-ray and radio images.  We finally draw our conclusions in Section \ref{conclusion}.

\section{Event selection and analysis} \label{selection}

\subsection{Selection criteria}

Our first goal is to observationally investigate the relationship between the starting frequency of groups of type III bursts and the spectral index of flare X-ray emission.  We used the RHESSI \citep{Lin_etal2002} satellite to obtain observations of hard X-ray spectra.  We used the Phoenix 2 \citep{Messmer_etal1999} spectrometer to obtain dynamic spectra of groups of type III radio bursts to obtain type III starting frequencies.  

Using RHESSI and Phoenix 2 posed some strict constraints.

\begin{itemize}
 
\item We selected events between 08:00 and 16:00 UT.

\item The type III starting frequencies must be within the 4~GHz to 100~MHz frequency range of Phoenix 2.  

\item All events must be outside the RHESSI night time and South Atlantic Anomaly.  

\item We want to select events that do not have complicated structures interfering with the signature of the type III burst.  
Therefore we considered events that had the `type III’ flag in the Phoenix 2 catalogue.

\end{itemize}

Using the RHESSI and Phoenix 2 catalogue of events they contained 14,174 X-ray flares and 1,046 groups of type III bursts.  There exists more type III bursts during this time but we only used events that had the ‘type III’ flag in the Phoenix 2 catalogue to satisfy the last constraint. These events do not contain other types of radio emission to ensures a clear identification of the type III burst starting frequencies.  We note that the groups of type III bursts refers to a collection of type III bursts that is caused by multiple electron beams accelerated during the course of a solar flare.  This is different from type III storms that can last hours to days and are related to an active region.  

We also desired images of our events.  We used RHESSI to obtain hard X-ray images and we used the Nan\c{c}ay Radioheliograph (NRH) \citep{KerdraonDelouis1997} for radio images.  The start and end observing times for the NRH vary throughout the year but are similar to the Phoenix 2 start and end times.  The start and end times of 08:00 and 16:00 UT cover the majority of events in Phoenix 2 and the NRH and it was simpler to keep static start and end times.  From the list of events derived from the RHESSI catalogue of flares and the type III catalogue of Phoenix 2 we applied the further conditions:

\begin{enumerate}
 
\item The X-ray peak time must occur within the duration of the group of type III radio bursts.

\item Time profile of the X-rays had to be impulsive.

\item X-ray emission is well detected above 25~keV so as to deduce HXR spectral indices without large error bars.

\item Groups of type III bursts must last 20 seconds or more.
\end{enumerate}

For condition 1 we used the time at which the peak of the X-ray flare was documented in the RHESSI flare list.  We wanted events that had a temporal co-existence of hard X-ray emissions and type III radio bursts.  

Condition 2 was implemented using information about the X-ray peak flux and duration from the RHESSI flare list.  We divided the peak flux by the duration to obtain a measure of the change in count rate (counts s$^{-2}$).  This condition was applied because an impulsive electron beam helps the generation of radio emission.  Flares that are extended in time or did not have high flux counts are undesirable.  We filtered the flares such that the peak flux / duration $> 0.1$.  This value of 0.1 as the threshold was arbitrary and was chosen because 1 reduced the number of flares too dramatically while 0.01 did not filter the flares enough.  Applying conditions 1 and 2, the list of events became 203 and 168, respectively.  

Condition 3 was imposed because we needed good spectral fits to the X-rays with small errors.  Condition 4 was imposed because we needed to have enough data points to correlate the type III starting frequencies with the X-ray spectral index in one single event.  We manually went through the list of events and finally selected 30 events which satisfied conditions 3 and 4 (see Table \ref{tab:Events}).

\subsection{Hard X-ray spectral index}

The hard X-ray spectral index was derived from RHESSI with a 2 second time integration (RHESSI half rotation).  A forward fit to the data was applied using a combination of a thermal and a simple power-law distribution using routines \textbf{vth} and \textbf{1pow} respectively from the OSPEX software.  The simple power-law fit was used over the routine \textbf{thick2} to obtain an approximation for the X-ray spectral index that contains less assumptions.  We also used the \textbf{albedo} routine to correct for reflected X-rays off the solar photosphere.  We assumed the electron distribution was isotropic \citep{DicksonKontar2013}.  We did not use the pulse-pileup correction because the majority of the flares had low count rates.  All the detectors were used for the spectral analysis except detectors 2 and 7.  Information on the RHESSI detectors can be found in \citet{Smith_etal2002} while a review on spectral analysis can be found in \citet{Kontar_etal2011}.  The \textbf{1pow} fitting routine gives the one sigma error on the hard X-ray spectral index.

\subsection{Radio starting frequency}\label{start_freq}

\begin{figure}
  \includegraphics[width=0.99\columnwidth]{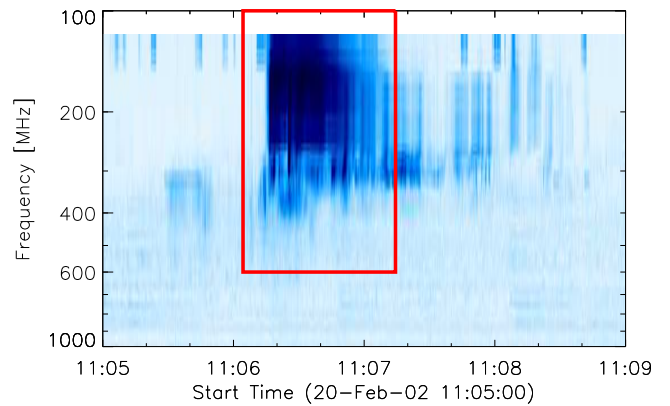}
\caption{Phoenix 2 dynamic spectrum of a group of type III radio bursts on the 20th February 2002 at 11:06 UT.  The group of bursts, enclosed by the red box, lasts for approximately 1 minute and starts around 600 MHz.}
\label{fig:spec}
\end{figure}

Using the Phoenix 2 radiospectrogram we obtained dynamic spectra of type III radio bursts.  The data was cleaned using the IDL SolarSoft routines \textbf{constbacksub}, \textbf{elimwrongchannels}, and the routine \textbf{LeeFilt}.  The time cadence for the spectrogram is 0.1 seconds.  An example spectrum is displayed in Figure \ref{fig:spec} for a flare on the 20th February 2002 (initially analysed by \citet{Vilmer_etal2002}).  The groups of type III bursts are generated from multiple electron beams accelerated over a minute, indicated by the almost vertical lines of radio emission especially noticeable between 200 and 400 MHz.  Unfortunately the 2-second time cadence for RHESSI constrains the radiospectrogram data to be compressed into 2 second bins.

To obtain the starting frequency from the radio spectrum we required a threshold value above the background signal that we deemed a type III burst.  After using the three algorithms mentioned above to clean up the dynamic spectra, their mean magnitude was zero.  We had to choose an arbitrary value above zero that was a significant signature of a type III burst.  This value was typically 2 or 3 but could be as high as 12 when there was a low level of decimetric emission at the same time as the groups of type III bursts (this occurred for only one event).  Moreover, certain channels could have larger noise that caused erroneous starting frequencies if the threshold was too low.  We used the lowest possible threshold for the starting frequency such that it observably (by eye) tracks the start of the type III bursts and was not compromised by high noise channels.  The width of the post-filtered frequency channels was used as the one sigma error on the starting frequency.

\subsection{Starting frequency vs X-ray spectral index}

\begin{figure*}
 \includegraphics[width=0.50\textwidth]{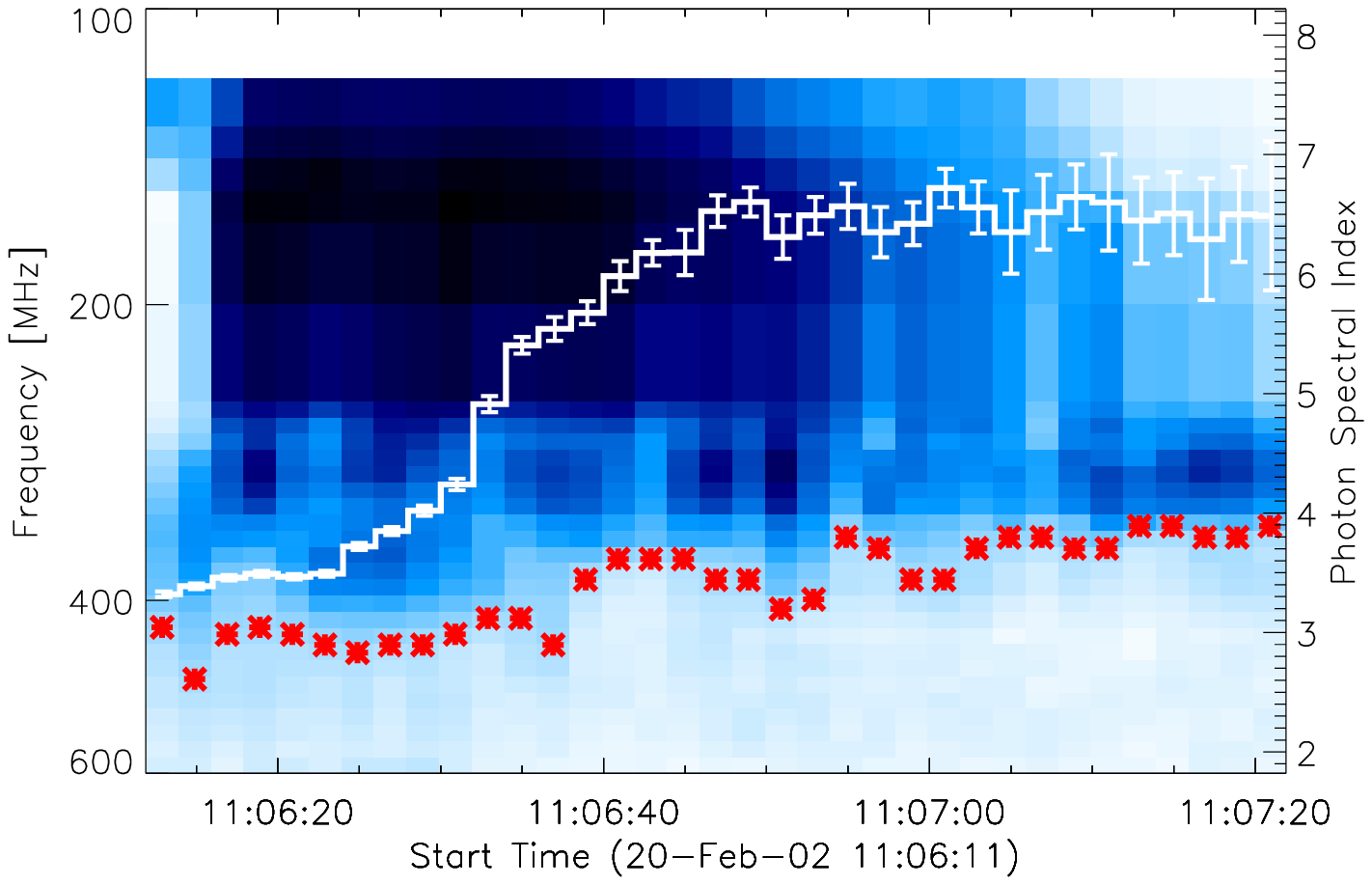}
 \includegraphics[width=0.50\textwidth]{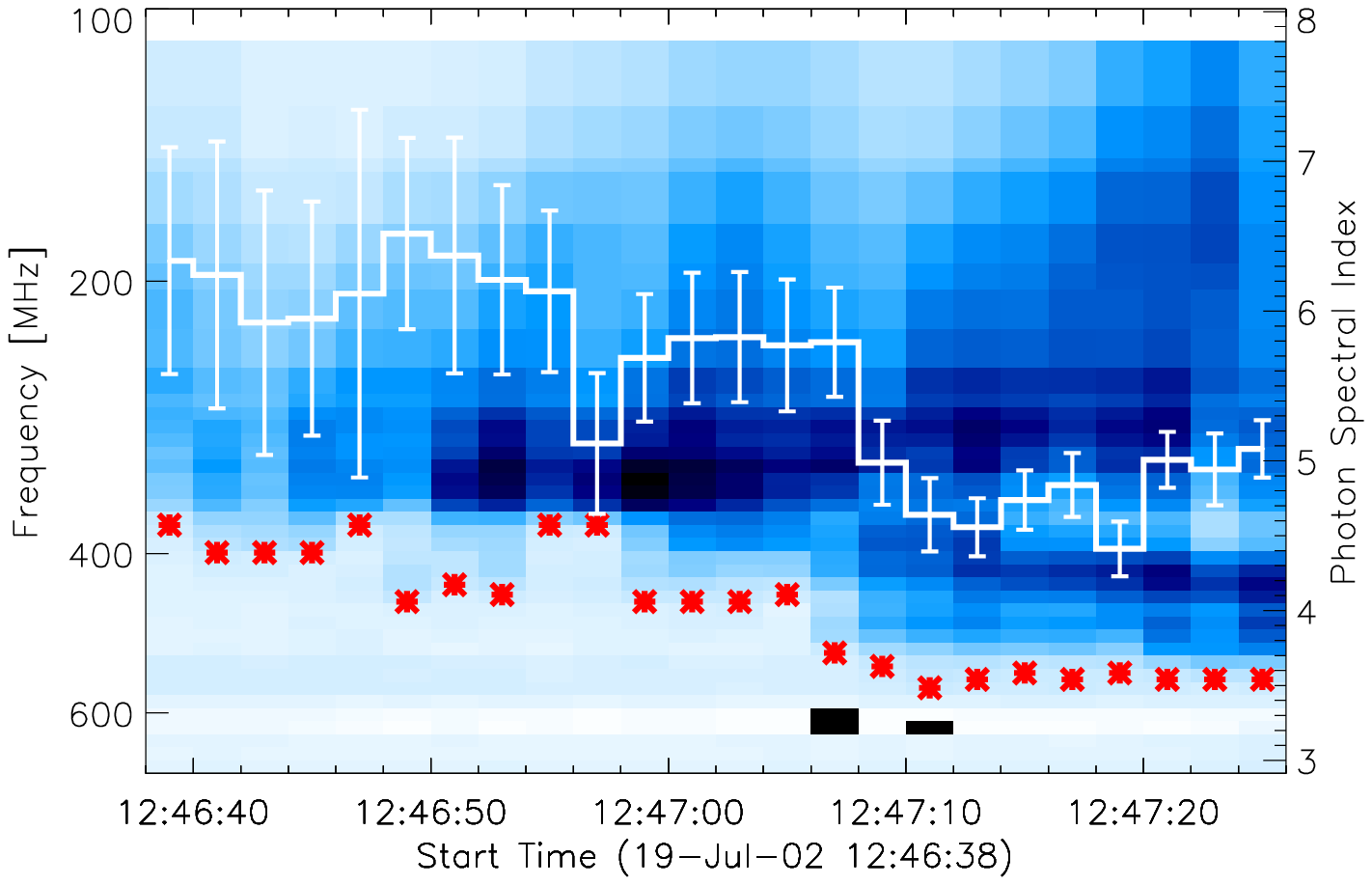}
 \includegraphics[width=0.50\textwidth]{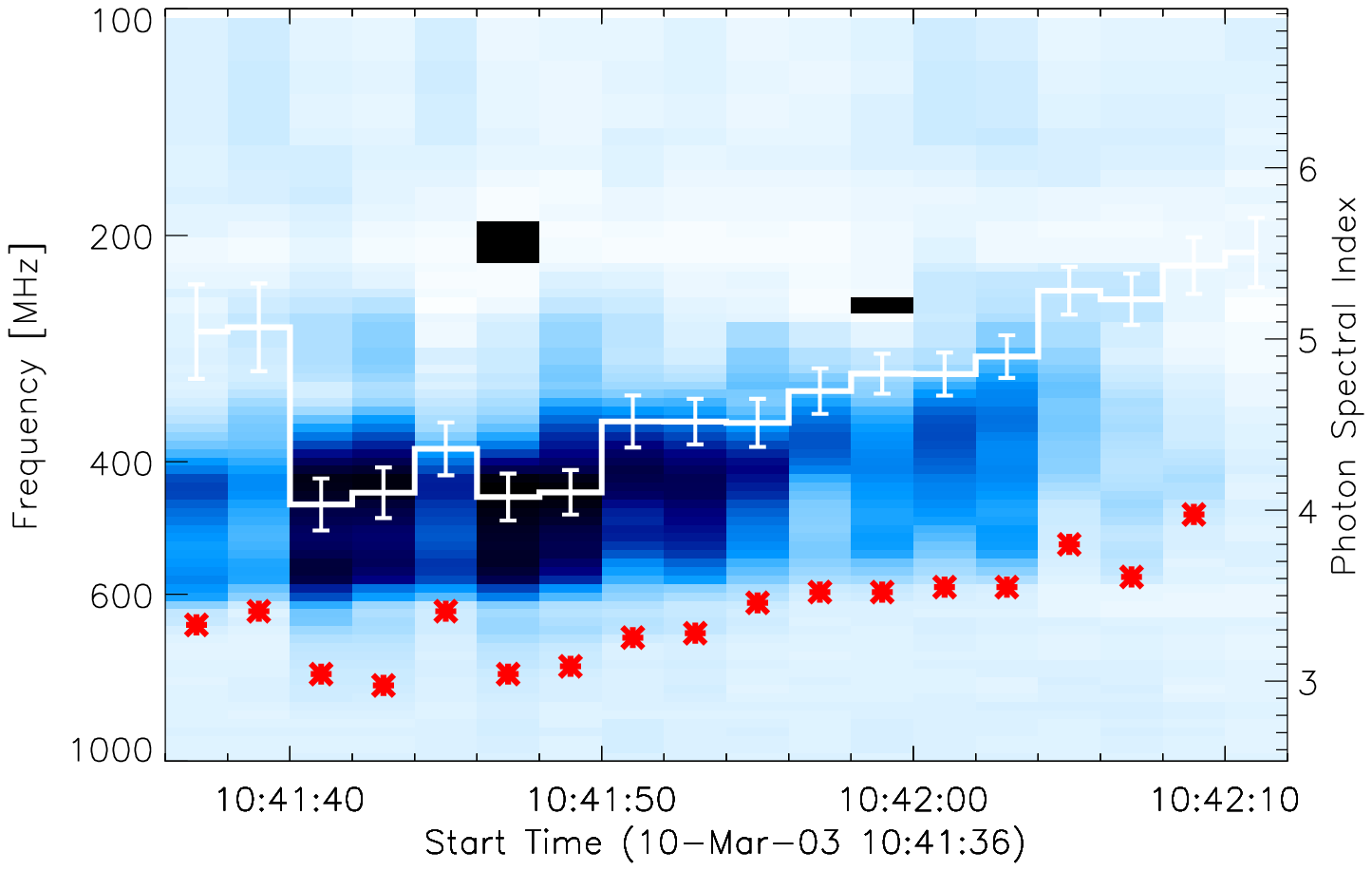}
 \includegraphics[width=0.50\textwidth]{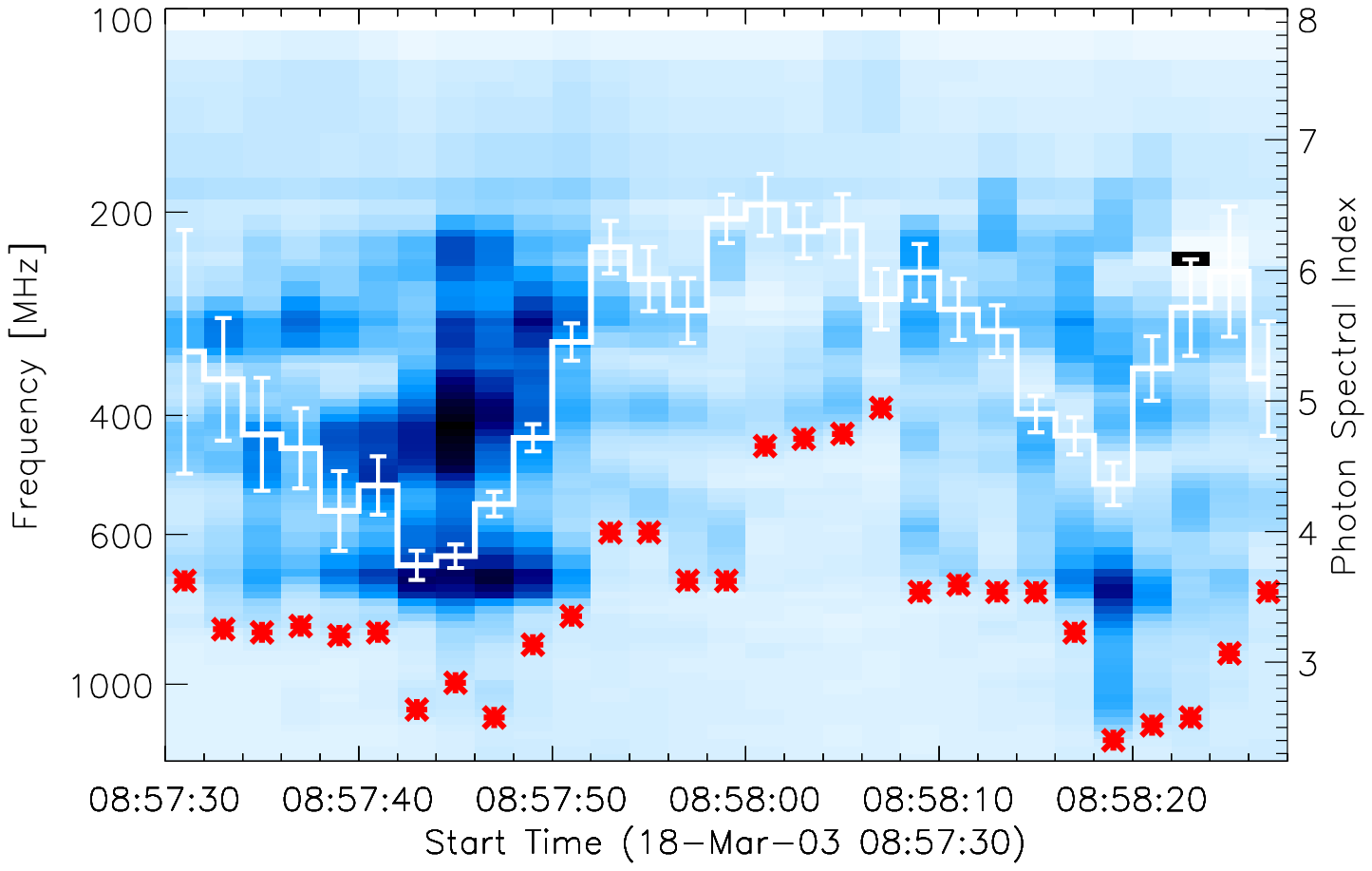}
 \includegraphics[width=0.50\textwidth]{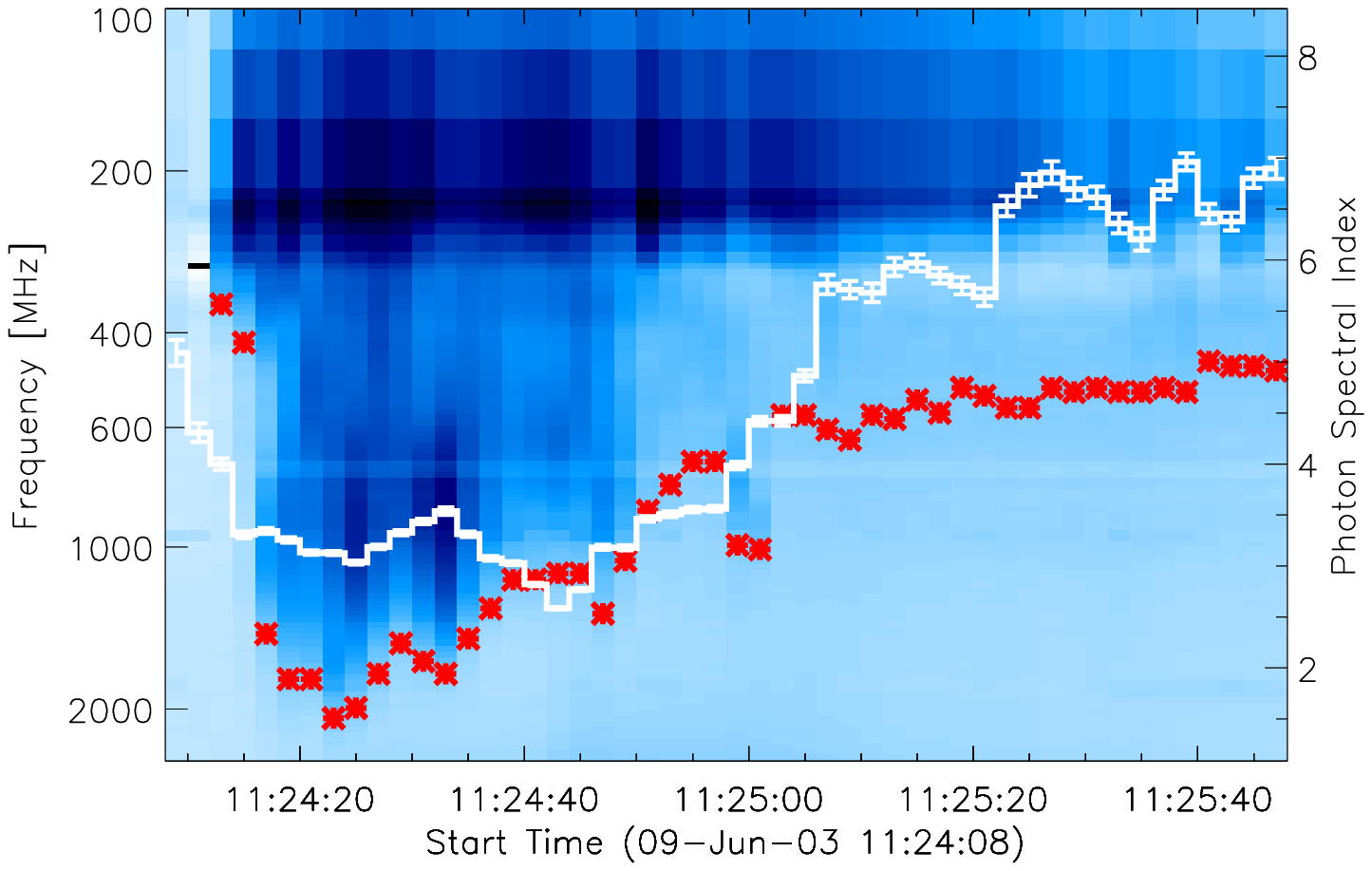}
 \includegraphics[width=0.50\textwidth]{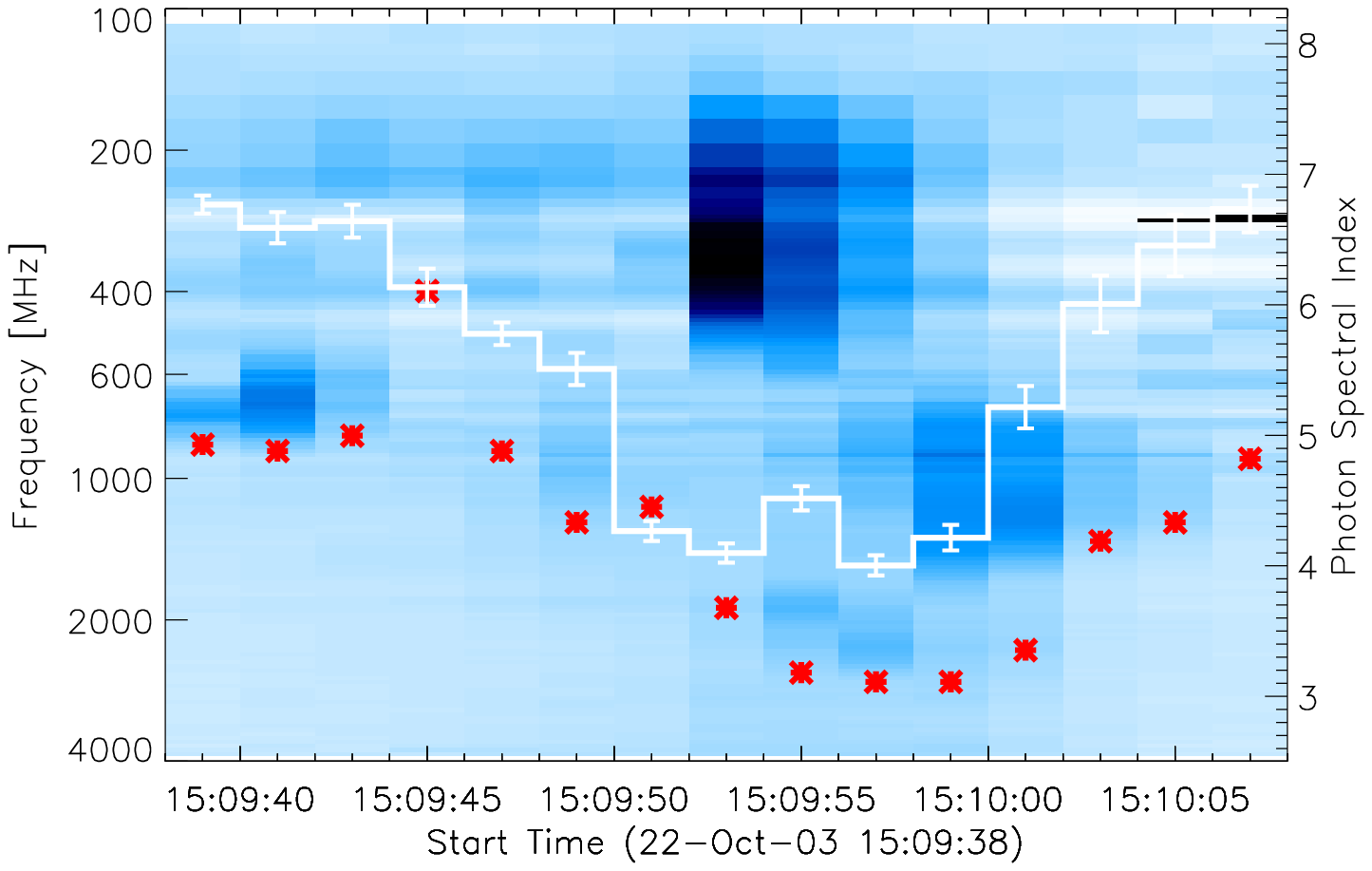}
\caption{Six examples of groups of Type III bursts using a time cadence of 2 seconds.  The background is radio observations from Phoenix 2.  Red stars are the starting frequencies found above the background.  White bars are the X-ray spectral index derived from RHESSI with the associated 1-sigma errors.  All six events show a link between the type III starting frequencies and the X-ray spectral indices.}
\label{fig:sfsi}
\end{figure*}

\begin{figure*}
 \includegraphics[width=0.50\textwidth]{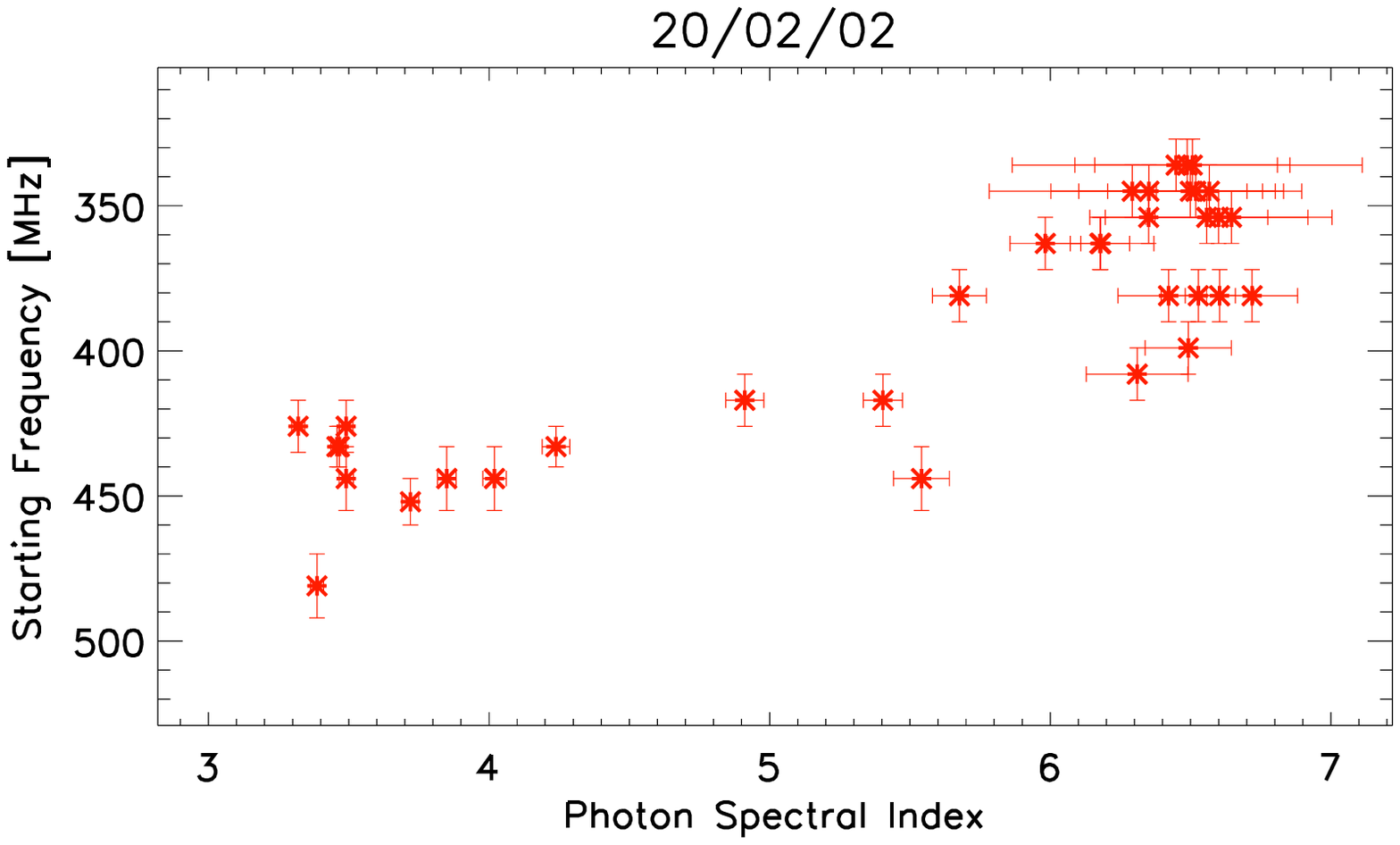}
 \includegraphics[width=0.50\textwidth]{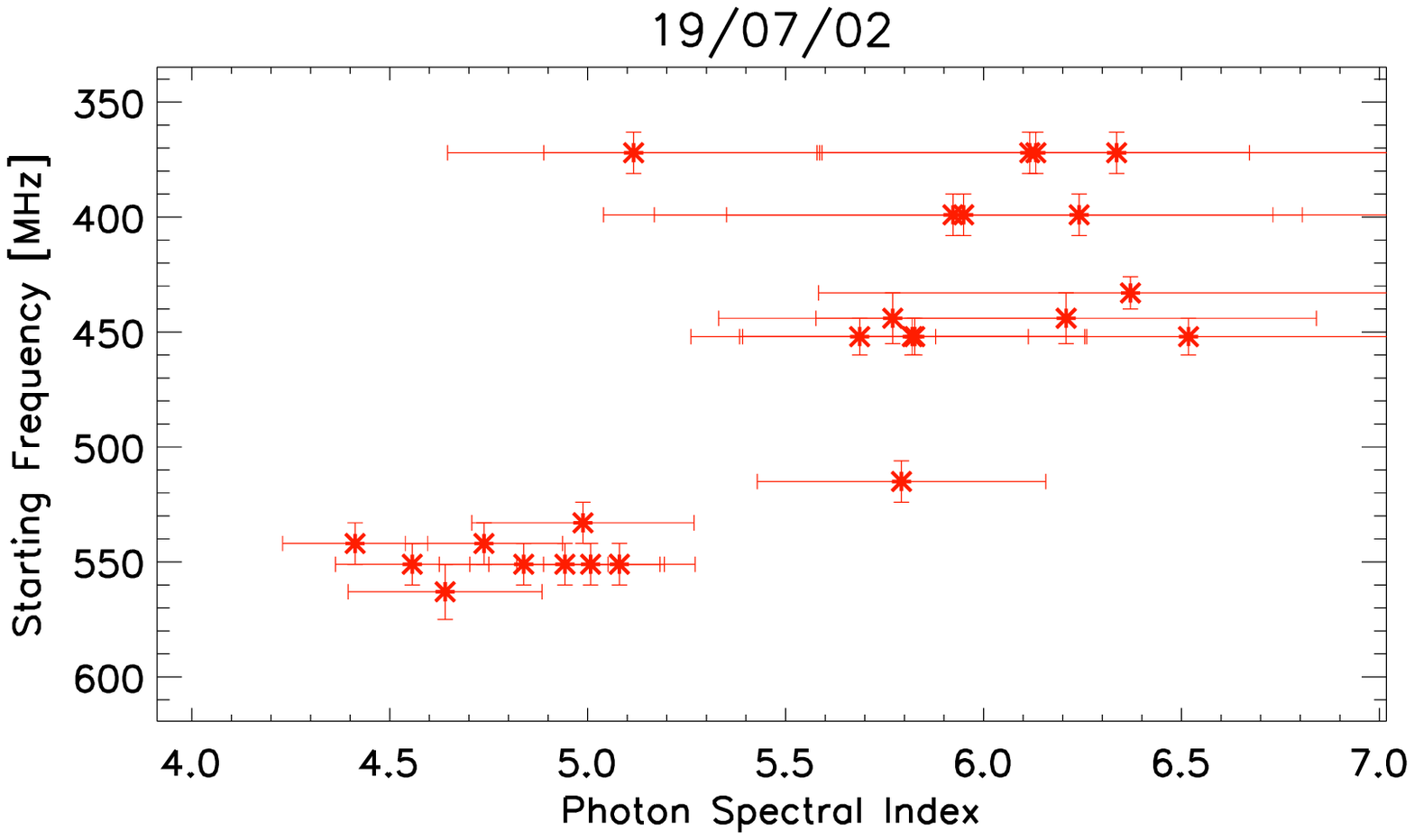}
 \includegraphics[width=0.50\textwidth]{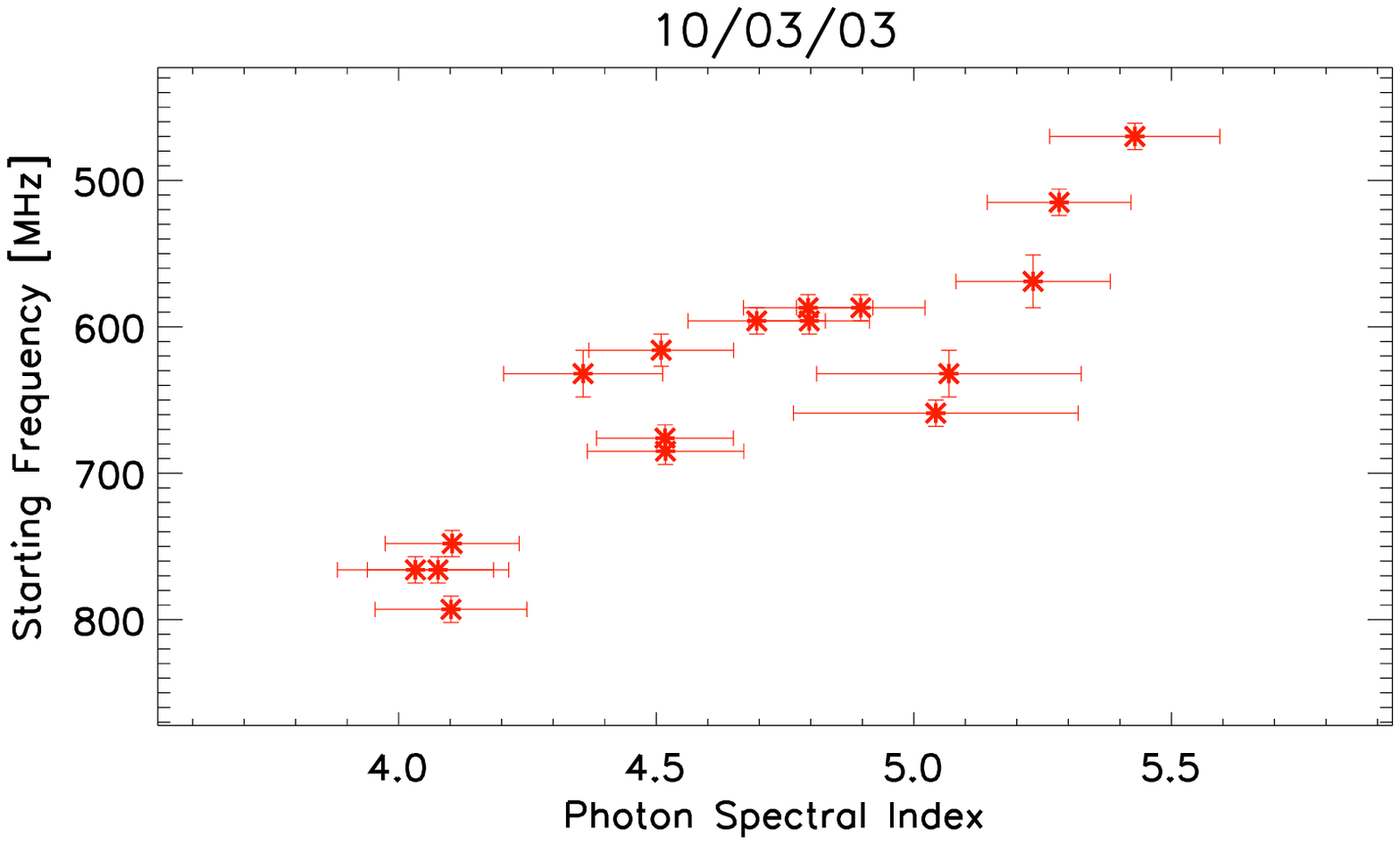}
 \includegraphics[width=0.50\textwidth]{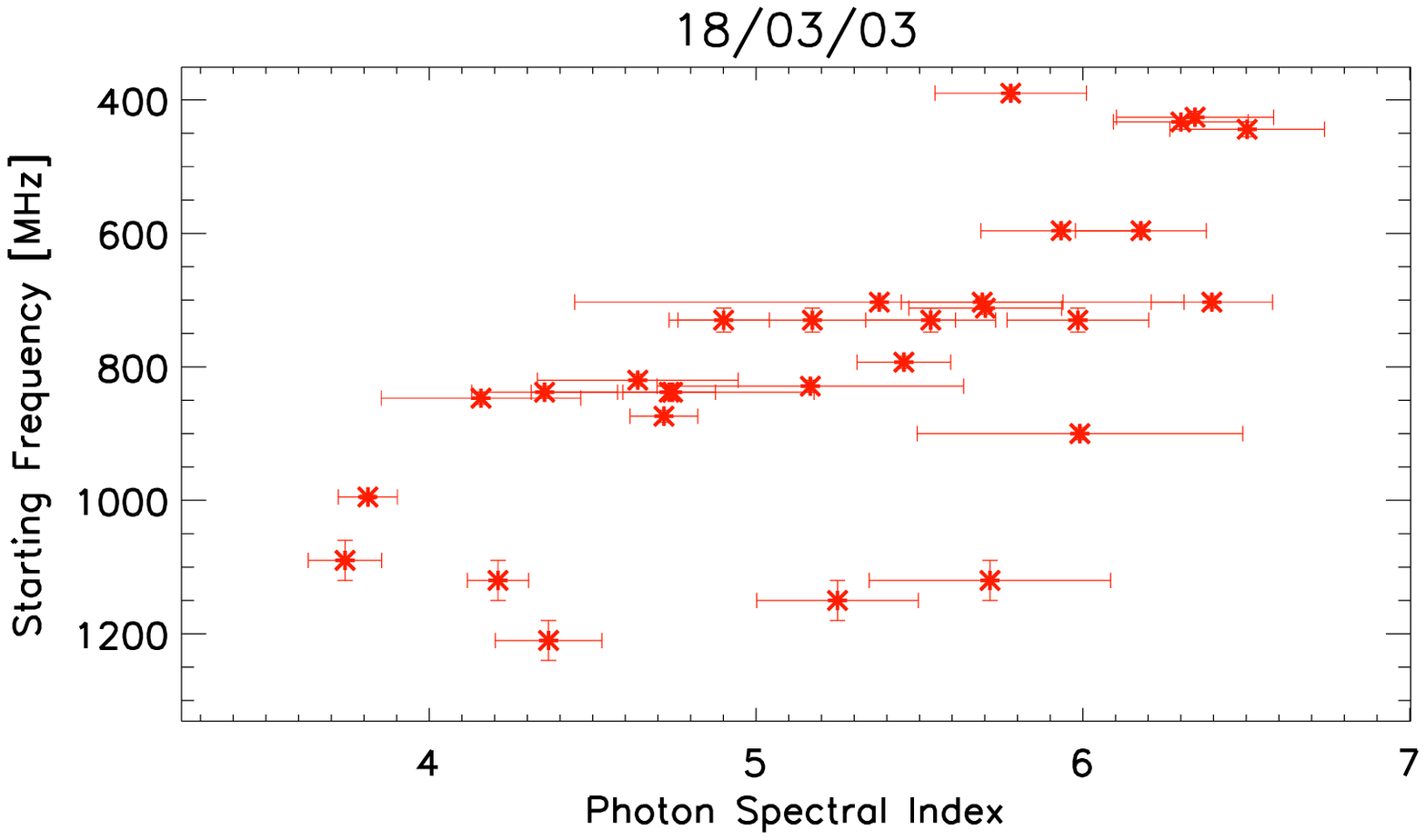}
 \includegraphics[width=0.50\textwidth]{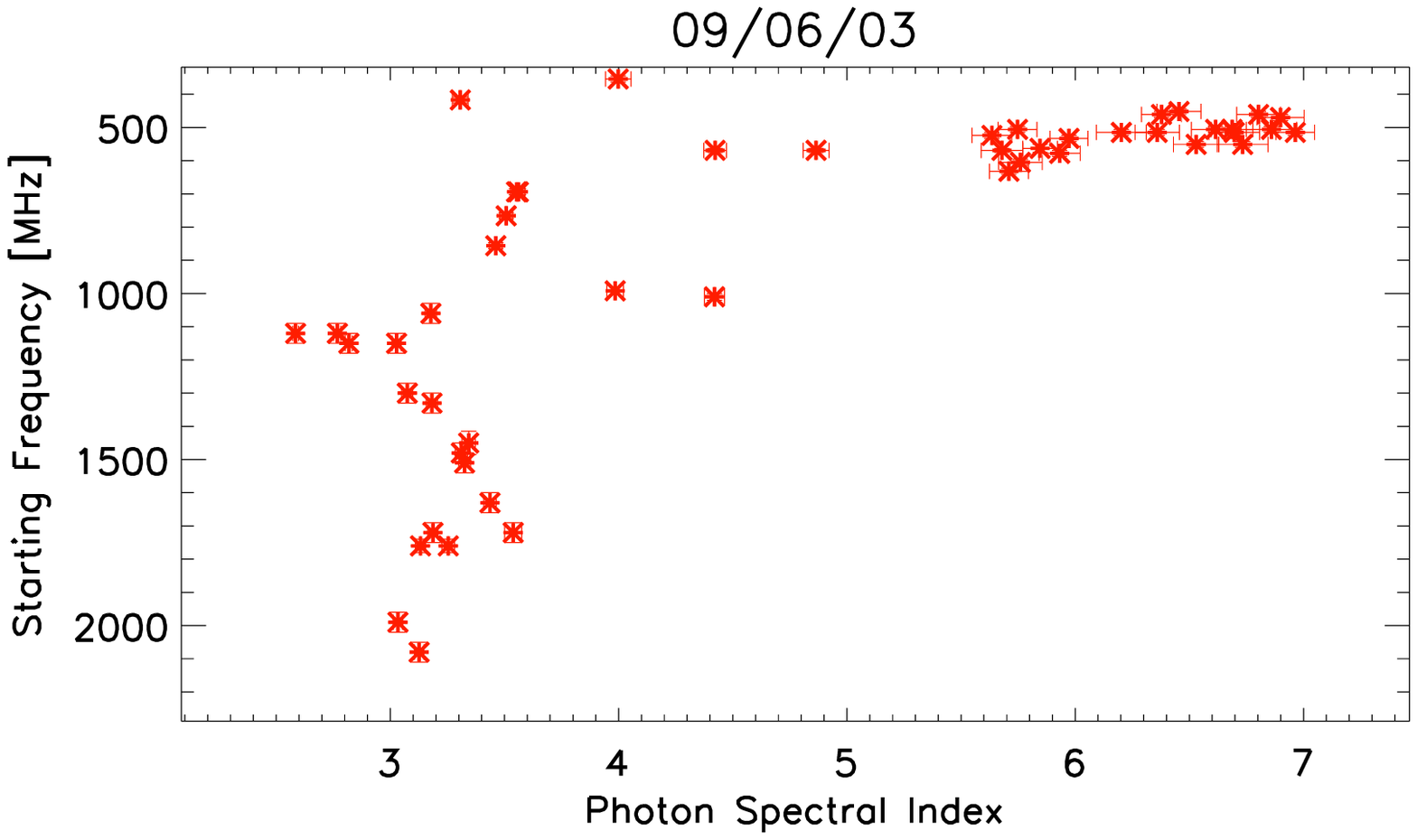}
 \includegraphics[width=0.50\textwidth]{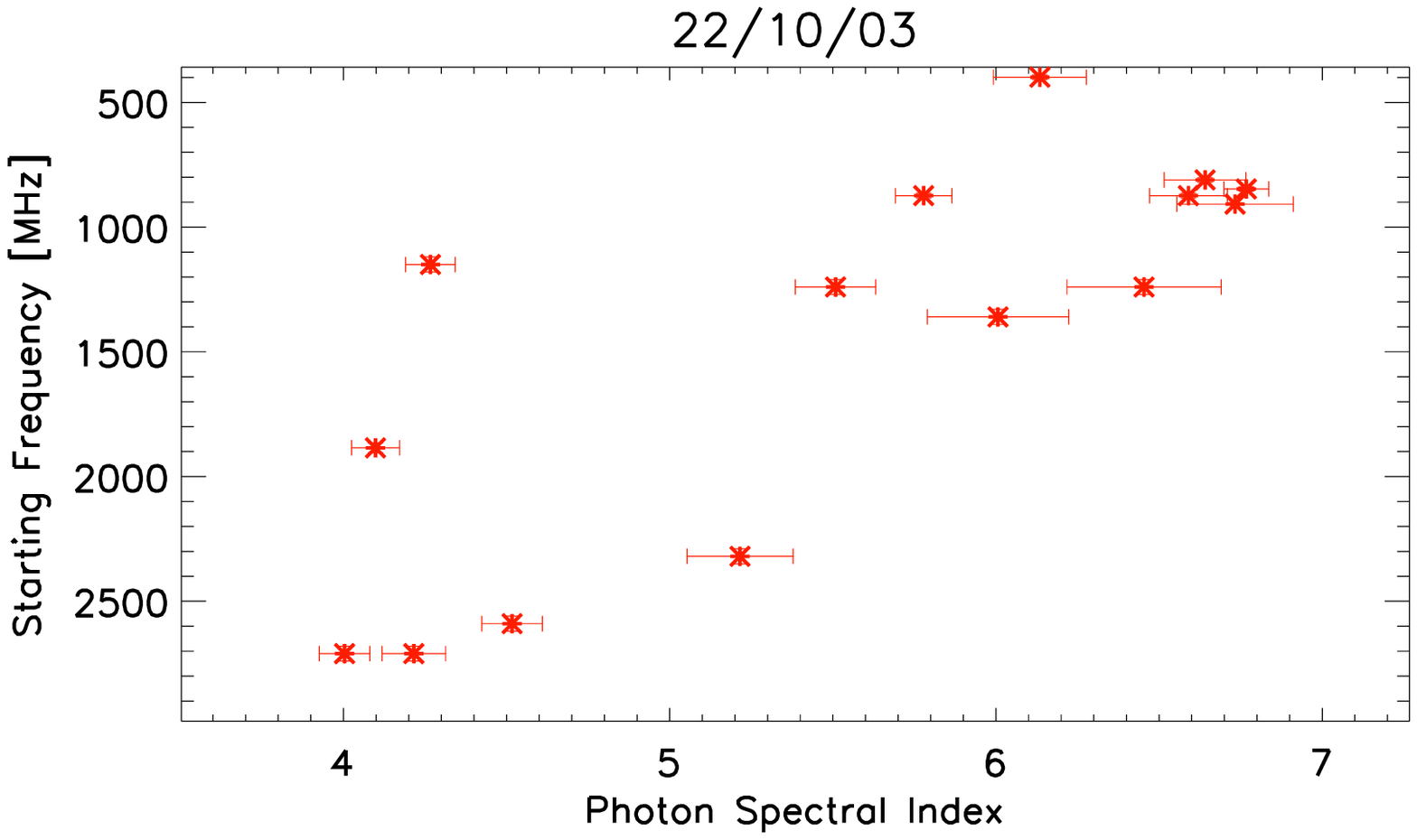}
\caption{Starting frequencies of the groups of type III bursts plotted against the X-ray spectral index for the events shown in Figure \ref{fig:sfsi}.  The one-sigma error bars are shown.}
\label{fig:fsi}
\end{figure*}

Some events show a connection between the type III starting frequencies and the hard X-ray spectral indices.  To illustrate this link we have plotted both quantities for the whole period of the groups of type III bursts for six different events (Figure \ref{fig:sfsi}).  The red stars can be observed to follow by eye the starting frequency of the radio bursts.  We can see that the spectral index follows the starting frequency; the spectral index is lower when the starting frequency is higher and vice-versa.  This trend is highlighted in the starting frequency vs. X-ray spectral index scatter plots of Figure \ref{fig:fsi} for all six events.  The selection of events shows how the starting frequencies change dramatically for some events and only slightly for others.  The highest starting frequencies in each event also varies from GHz down to 500 MHz.  Figure \ref{fig:sfsi} also shows how blurred the radio dynamic spectrum appears when it is averaged into 2 second time bins.

We also observed that the groups of type III bursts starting frequencies tend to exhibit a low-high-low pattern.  This is particularly evident on the 18th March 2003 in Figure \ref{fig:sfsi}.  For this event we observe two instances of the X-ray spectral indices going from soft to hard and then to soft again.  The corresponding type III starting frequencies go from low to high and then to low again, or exhibit a low-high-low pattern.  Our selections of start and end times cut off part of this pattern for some of the other events in Figure \ref{fig:sfsi}.  Given our expectancy of a correlation between the electron beam spectral index and the starting height of the type III bursts \citep[e.g.][]{Reid_etal2011} it is not surprising that we observe this low-high-low pattern for the starting frequencies.



\begin{center}
\begin{table*}
\centering
\begin{tabular}{ c  c  c  c  c  c  c }

\hline\hline

 Event Date &  Start Time & End Time & Starting Frequency   & Radio SF vs X-ray SI    & Starting Height vs Electron SI & Included in \\
 		    &  			  & 		 &  Threshold 			& Correlation Coefficient & Correlation Coefficient        & this Study \\ \hline

 14/02/02  &  11:03:30 & 11:04:12 &  3   &   -0.24  &   0.18  &  \\
 14/02/02  &  13:51:38 & 13:52:30 &  2   &   -0.60  &   0.63  &  $\circ$    \\
 20/02/02  &  11:06:11 & 11:07:22 &  2   &   -0.86  &   0.85  &  $\bullet$  \\
 28/02/02  &  14:14:12 & 14:14:52 &  6   &   0.08   &  -0.02  &  \\
 15/04/02  &  08:51:20 & 08:53:12 &  2   &   -0.44  &   0.43  &  \\
 20/04/02  &  13:34:48 & 13:35:18 &  3   &   -0.76  &   0.77  &  $\bullet$  \\
 02/06/02  &  11:44:16 & 11:45:18 &  6   &   -0.86  &   0.84  &  $\bullet$  \\
 26/06/02  &  15:21:42 & 15:22:14 &  2   &   -0.19  &   0.20  &  \\
 19/07/02  &  12:46:38 & 12:47:26 &  1   &   -0.79  &   0.77  &  $\bullet$  \\
 20/08/02  &  08:25:18 & 08:26:10 &  7   &   0.18   &  -0.20  &  \\
 23/08/02  &  11:05:48 & 11:06:34 &  2   &   -0.27  &   0.31  &  \\
 10/09/02  &  14:52:30 & 14:53:30 &  12  &   -0.77  &   0.77  &  $\bullet$  \\
 14/09/02  &  12:22:20 & 12:23:06 &  2   &   -0.81  &   0.81  &  $\bullet$  \\
 17/09/02  &  10:37:40 & 10:40:20 &  2   &   -0.66  &   0.63  &  $\circ$    \\
 29/09/02  &  14:48:28 & 14:49:12 &  2   &   -0.58  &   0.57  &  $\circ$    \\
 10/03/03  &  10:41:36 & 10:42:12 &  1   &   -0.89  &   0.89  &  $\bullet$  \\
 18/03/03  &  08:57:30 & 08:58:28 &  2   &   -0.63  &   0.85  &  $\bullet$  \\
 26/04/03  &  08:05:00 & 08:07:04 &  4   &   -0.33  &   0.40  &  \\
 09/06/03  &  11:24:08 & 11:25:48 &  5   &   -0.74  &   0.78  &  $\circ$    \\
 12/06/03  &  11:41:20 & 11:43:06 &  2   &   -0.74  &   0.67  &  $\bullet$  \\
 12/06/03  &  14:00:26 & 14:00:46 &  5   &   -0.50  &   0.47  &  \\
 22/06/03  &  09:47:16 & 09:48:48 &  2   &   -0.33  &   0.34  &  \\
 09/07/03  &  13:20:26 & 13:22:44 &  2   &   -0.50  &   0.50  &  \\
 07/08/03  &  13:12:02 & 13:12:42 &  2   &   -0.60  &   0.67  &  $\circ$    \\
 27/09/03  &  11:57:04 & 11:59:40 &  2   &   -0.11  &   0.11  &  \\
 30/09/03  &  08:48:40 & 08:49:16 &  2   &   -0.61  &   0.67  &  $\circ$    \\
 22/10/03  &  15:09:38 & 15:10:08 &  2   &   -0.78  &   0.74  &  $\circ$    \\
 28/05/04  &  08:13:52 & 08:14:12 &  2   &    0.17  &  -0.20  &  \\
 25/07/04  &  13:22:32 & 13:23:16 &  3   &   -0.69  &   0.72  &  $\bullet$  \\
 09/05/05  &  13:57:18 & 13:57:42 &  3   &   -0.41  &   0.41  &  \\

\hline
\end{tabular}
\vspace{20pt}
\caption{A chronological list of the events analysed in the study.  Included is the start time in UT, the threshold above the background used for the starting frequency, the correlation coefficient found between the starting frequencies of the groups of type III bursts and the hard X-ray spectral indices, and the correlation coefficient between the derived starting height of groups of type III bursts and the derived electron velocity spectral index.  The last column indicates the high magnitude correlation coefficient events, with a full circle if the event was able to estimate acceleration region parameters or an empty circle if it was not.}
\label{tab:Events}
\end{table*}
\end{center}

\subsection{Event List}

We now consider all of the 30 events in our study (see Section \ref{selection}).  Table \ref{tab:Events} shows a list of all the event dates.  The start time and end time in UT is also shown along with the threshold value above the background signal that was used to find the radio starting frequencies.  The start and end time for each event indicates the earliest and latest time respectively that there was enough radio emission for a starting frequency to be found and that an X-ray spectral index was found without large error values that was not too soft (usually below 7).

For each event we computed the Pearson correlation coefficient \citep[e.g.][]{NumericalRecipies1992} (CC) between the starting frequencies of the groups of type III bursts and the X-ray spectral spectral indices.  This is indicated in the fifth column of Table \ref{tab:Events}.  We find that 17 out of the 30 events show a good anti-correlation (we deemed the CC magnitude 0.55 and greater to be good) between radio starting frequency and X-ray spectral index over the entire event (indicated by circles in the last column of Table \ref{tab:Events}).  We might not expect such a simple correlation given the complicated non-linear nature of coherent radio emission compared to the Bremsstrahlung process responsible for hard X-rays.  Moreover, some of the events that do not show a good anti-correlation over the entire event still show signatures of the starting frequencies following the spectral index in smaller time fragments (see Section \ref{discussion} for more discussion).  The second last column will be discussed in the next section.

The results of this systematic study finally show that for $50\%$ of the events, a good anti-correlation is observed between the starting frequencies and HXR. This trend reveals that for half of the events the simple cartoon of Figure \ref{fig:overview} is a good representation of the relative morphology of the X-ray and radio sources. For those events, a low-high-low trend of the starting frequencies is observed in relation to the soft-hard-soft evolution of the HXR spectrum during flares.

\section{Electron beam properties} \label{electron_beam}

We are now going to use some standard assumptions to deduce electron beam properties from the X-ray spectral index and the radio starting frequency.  Spectral indices of the electron beams will be deduced from the HXR spectral indices.  From starting frequencies of the type III radio bursts we deduce the height in the solar atmosphere that the electron beam became unstable to Langmuir waves that subsequently caused the emission of the type III radio bursts.

\subsection{Electron velocity spectral index}\label{spec_index}

To obtain the velocity spectral index of the electron beam we made the assumption that the electron beams that travel both upwards into the high corona and downwards into the chromosphere are produced in the coronal acceleration region (see Section \ref{intro} and Figure \ref{fig:overview}). The upward and downward propagating electron beams are furthermore supposed to have the same energy spectrum - the energy distribution of accelerated electrons is independent of initial propagation direction.  Using this assumption we obtained the upward velocity spectral index of the electron beam using the observed X-ray spectral index.


To deduce the \emph{electron velocity spectral index} from the X-ray spectral index, we have used the thick-target model \citep[e.g.][]{Brown1971, Brown_etal2006}.  The electron velocity spectral index $\alpha$ is obtained from the hard X-ray spectral index $\gamma$ using the thick target model with the relation $\alpha = 2(\gamma+1)$.  The one-sigma errors on the electron velocity spectral index are found by doubling the one-sigma errors on the hard X-ray spectral index.

\subsection{Starting height}\label{star_height}

To deduce from the starting frequencies the starting heights in the solar atmosphere at which the electron beam became unstable to Langmuir wave growth that subsequently caused radio waves we must assume a background electron density model. The density related to the radio starting frequency combined with the density model allows us to deduce the starting height.  We have assumed second harmonic emission as the emission mechanism from Langmuir waves to electromagnetic emission. One-sigma errors for the starting heights were calculated using the distance that corresponded to the one-sigma errors on the starting frequencies.

In a previous study \citep{Reid_etal2011} we used the exponential density model presented in \citet{Paesold_etal2001}.  Whilst being adequate, the model cannot account for plasma frequencies above 700 MHz (second harmonic emission of 1400 MHz), being below the surface of the Sun.  Type III bursts are known to travel in over-dense loops or coronal structures \citep{Wild_etal1959} which explains observed high frequencies (2~GHz), and hence high densities ($1.24\times10^{10}~\rm{cm}^{-3}$) of the ambient coronal electrons (assuming harmonic emission).  Simple scaling (e.g. 2$\times$, 5$\times$) of the exponential density model or other common models like the Baumbach-Allen \citep{Allen1947} and Saito \citep{Saito_etal1970} models can replicate higher densities but scaling does not affect the density gradient and can lead to improbably high densities in the high corona.  The exponential density model takes the form
\begin{equation}\label{exp_dens}
n(r) = A\exp(-r/r_n) 
\end{equation}
where $r$ is the distance taken from the centre of the Sun and $r_n$ is the characteristic scale height of the density model.  The normalisation constant $A$ is found from the reference values in \citet{Paesold_etal2001} of $n_0=3.36\times10^{8}~\rm{cm}^{-3}$ and $h_0=2.21\times10^{10}~\rm{cm}$ via $A = n_0\exp((h_0+R_{\bigodot})/r_n)
$.  Previously \citep{Reid_etal2011} $r_n$ had the value of $7.5\times10^9~\rm{cm}$ given in \citet{Paesold_etal2001}.  We have used a value of $r_n=3.16\times 10^{9}~\rm{cm}$ ($10^{9.5}~\rm{cm}$) which is the value derived from 10,000 type III radio burst observations \citep{Saint-Hilaire_etal2013}.  The lower value of $r_n$ leads to a higher density gradient $dn/dr$ and can account for higher densities in the low corona.

\subsection{Starting height vs electron spectral index}

We now examine whether the deduced starting heights correlate with the electron beam spectral indices.  The second last column of Table \ref{tab:Events} shows the values of the correlation coefficients between the starting heights of the type III radio bursts and the electron velocity spectral indices.  Again we find that 17 of the 30 events show a good correlation (CC magnitude  0.55 or greater), the same events as before (indicated by circles in Table \ref{tab:Events}).  Due to the inclusion of a electron density model that is not linear, this result was not necessarily predictable.  However, it was physically motivated by the theoretical connection between electron beam starting height and the distance an electron beam will travel before it becomes unstable to Langmuir wave growth.

\section{Inference of acceleration region parameters} \label{properties}

In this section we follow the procedure previously used in \citet{Reid_etal2011} to infer the height and vertical extent of electron acceleration regions.  We will first briefly describe this model and then use the parameters derived in Section \ref{electron_beam} with the model to deduce some of the properties of the electron acceleration regions.



\subsection{Electron beam dynamics} \label{equations}

We assume here (see Figure \ref{fig:overview}) that the accelerated electron beam starts at position $h_{acc}$ and becomes unstable to Langmuir wave growth at position $h_{typeIII}$ (the starting height of the type III burst), travelling distance $\Delta r=h_{typeIII}-h_{acc}$.  We find (see Appendix \ref{app1} for the derivation) that we can approximate $\Delta r=d\alpha$, where $d$ is the vertical spatial size of the electron beam and $\alpha$ is the spectral index of the electron beam in velocity space.  We thus have the following relationship
\begin{eqnarray}
h_{typeIII}=d\alpha + h_{acc},
\label{eq_mx_c}
\end{eqnarray}
Using the two observationally derived parameters of Section \ref{electron_beam}, namely $h_{typeIII}$ and $\alpha$, we will fit a straight line to each event and deduce $h_{acc}$ and $d$ from the gradient and intercept of each fit.

\subsection{Parameters of the acceleration regions}

\begin{figure*}
 \includegraphics[width=0.50\textwidth]{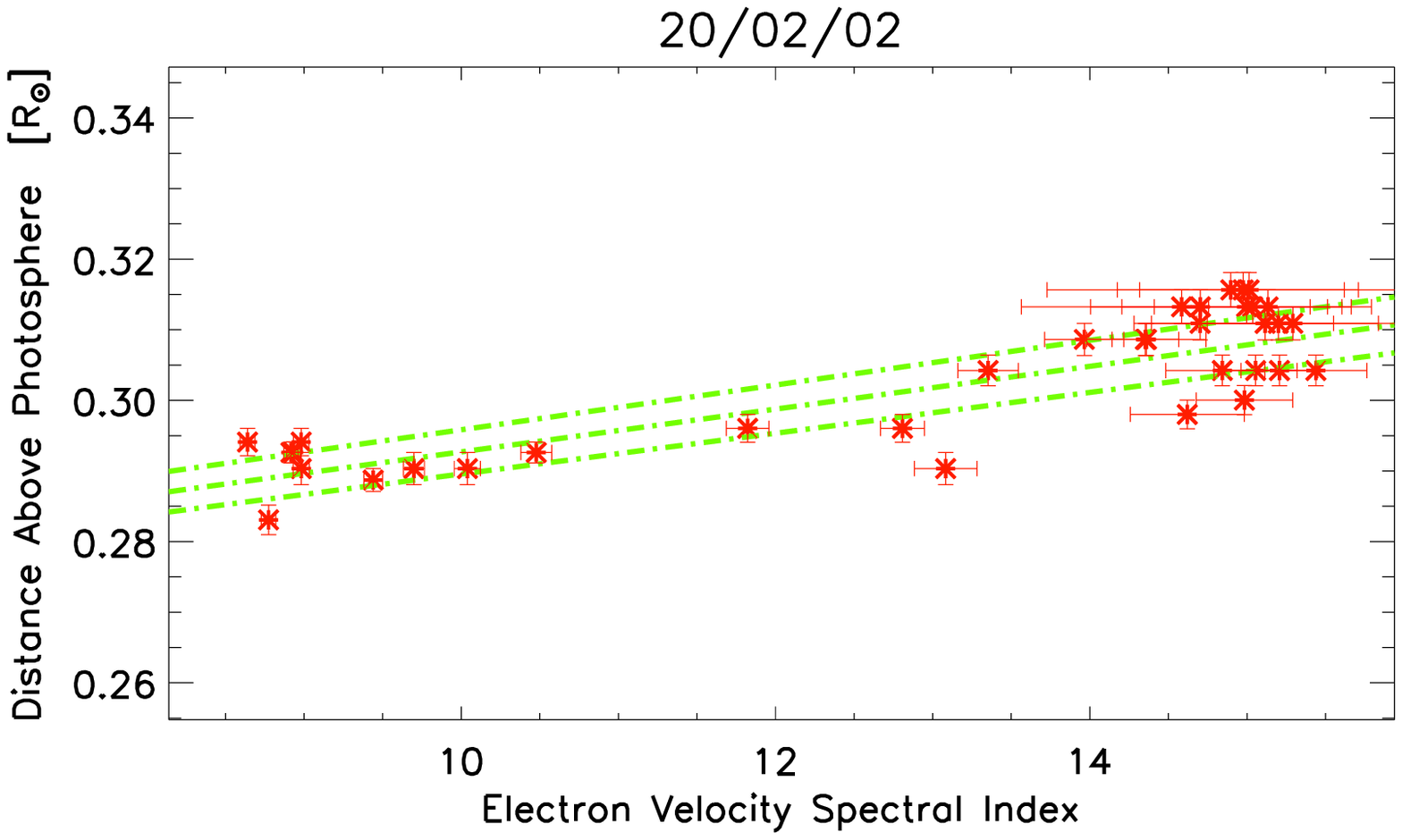}
 \includegraphics[width=0.50\textwidth]{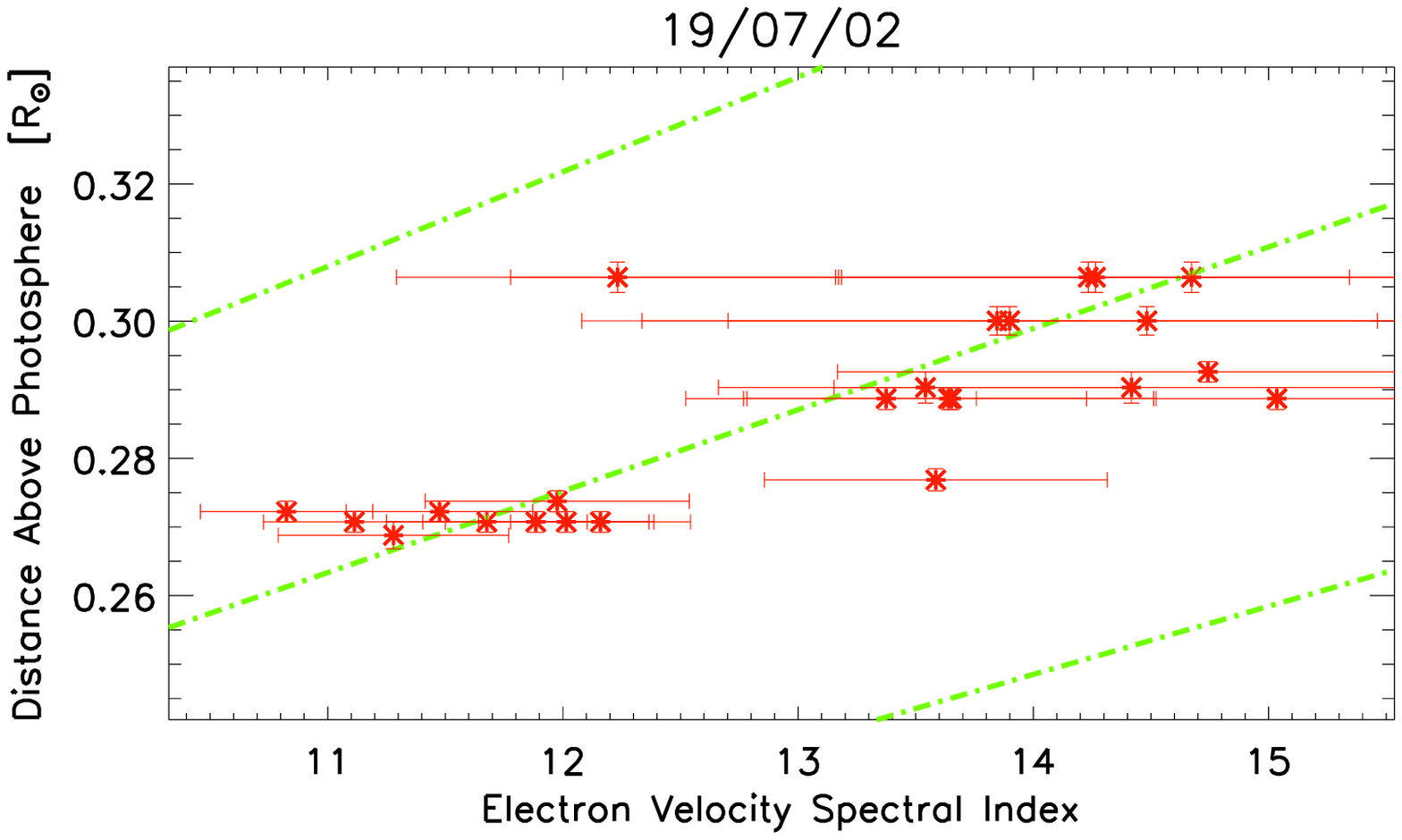}
 \includegraphics[width=0.50\textwidth]{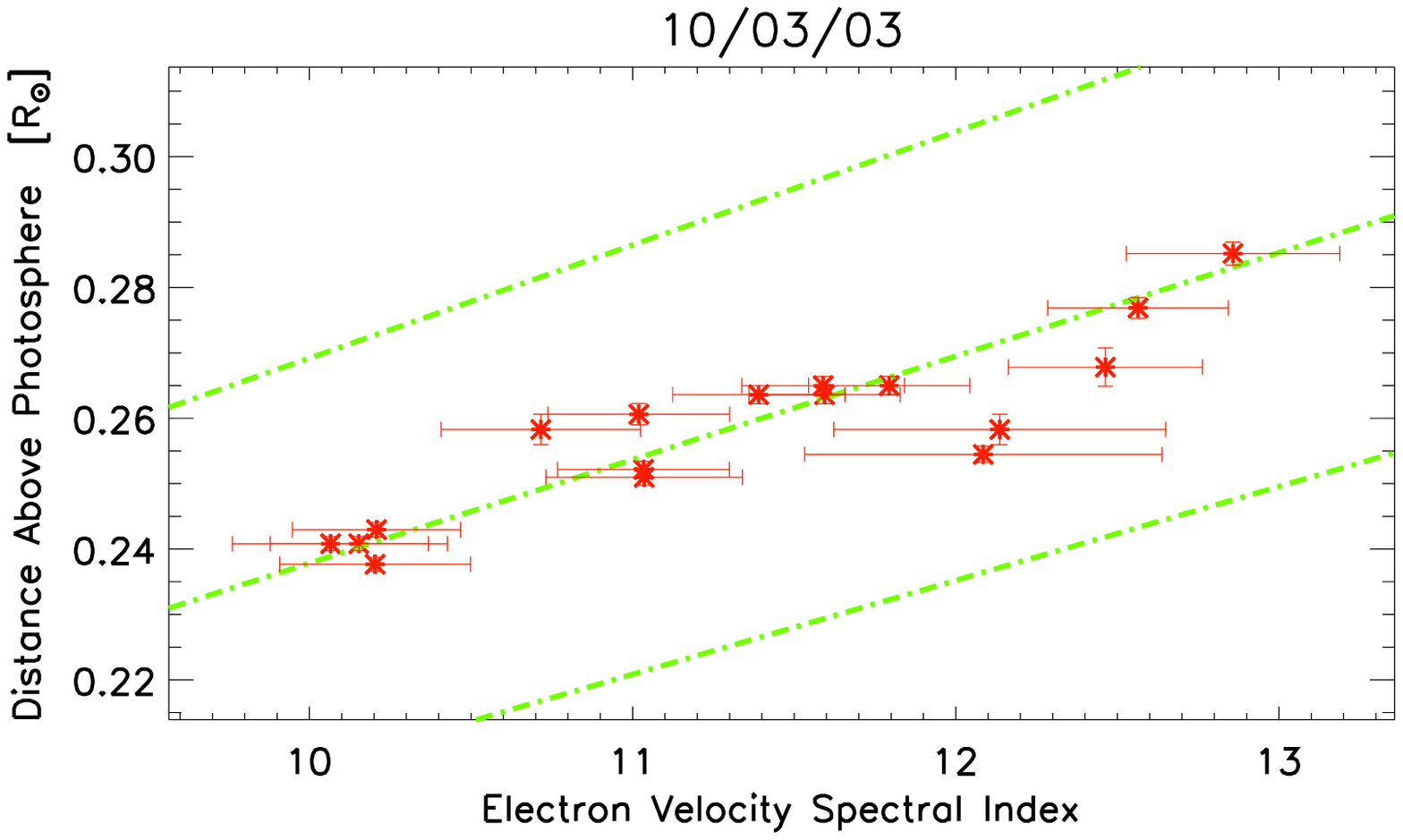}
 \includegraphics[width=0.50\textwidth]{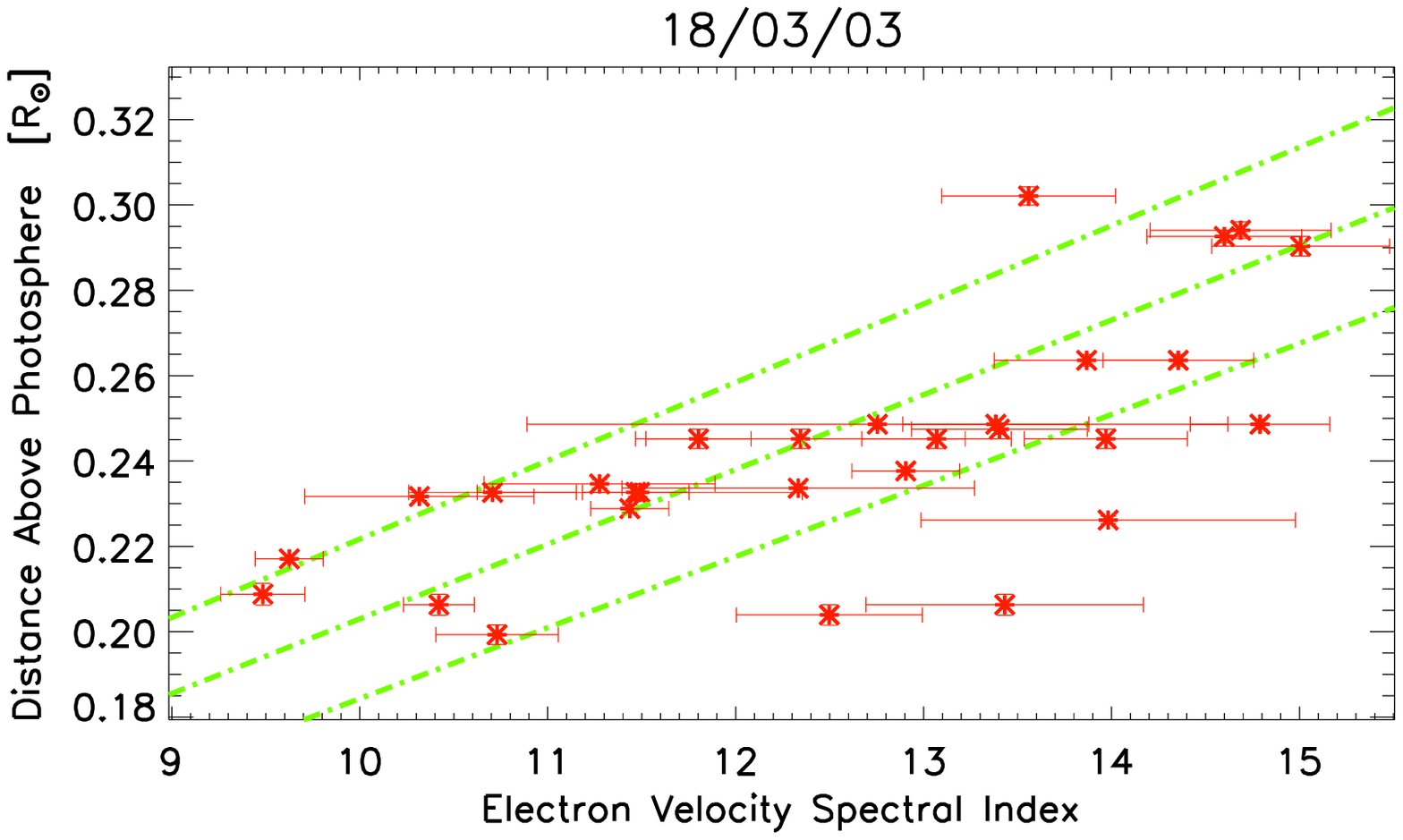}
 \includegraphics[width=0.50\textwidth]{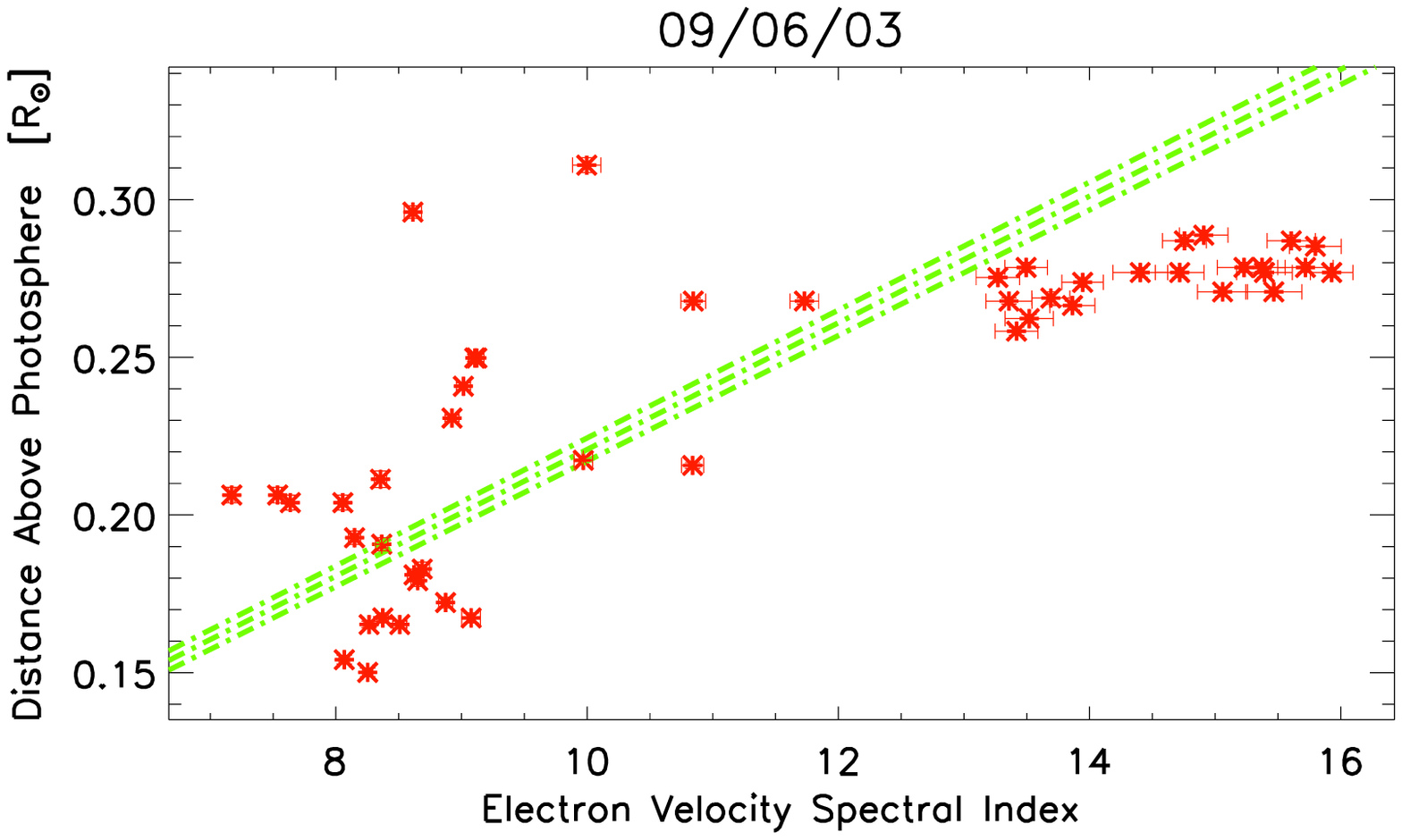}
 \includegraphics[width=0.50\textwidth]{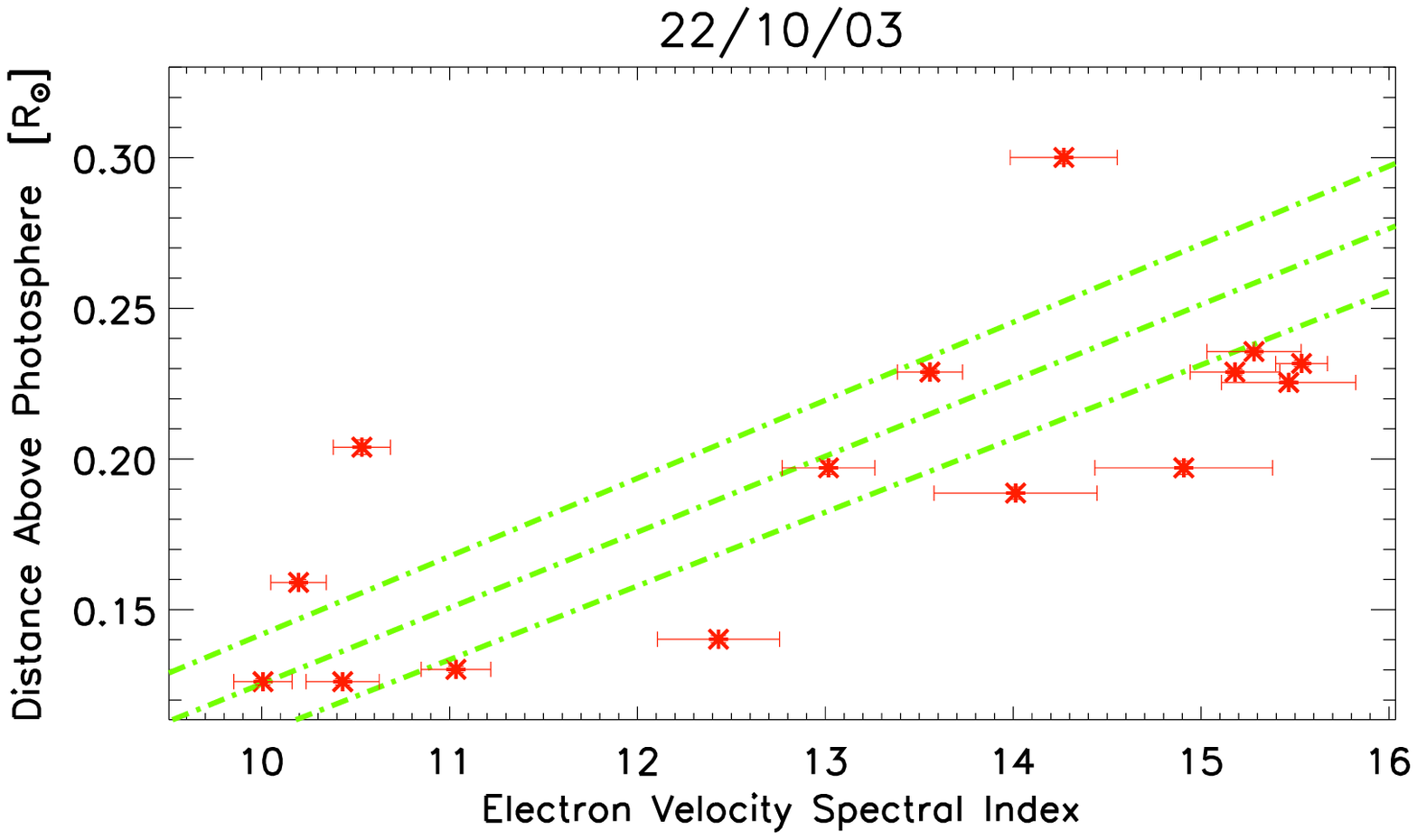}
\caption{Starting heights of the groups of type III bursts plotted against the electron velocity spectral index for the events shown in Figure \ref{fig:sfsi}.  The central dashed line shows the linear fit to the data with the outer dashed lines showing the one-sigma extremes of the fit.}
\label{fig:dsi}
\end{figure*}

Using the assumptions detailed in Sections \ref{spec_index} and \ref{star_height}, Figure \ref{fig:dsi} shows the starting heights of the groups of type III bursts of Figure \ref{fig:sfsi} as a function of electron velocity spectral index.  We have used the relation $h_{typeIII} = d\alpha + h_{acc}$ of Equation (\ref{eq_mx_c}) to fit the data with a straight line.  The central green dashed line in Figure \ref{fig:dsi} is a linear fit to the data using the routine \textbf{mpfitexy} \citep{Williams_etal2010} which depends on the \textbf{mpfit} package \citep{Markwardt2009}.  The fit takes into account the errors on both the electron velocity spectral index and the starting height.  The outer green dashed lines in Figure \ref{fig:dsi} show the extremes of the fit.  The fit to the event of the 20th February 2002 was particularly well constrained while the fit to the event of the 19th July 2002 is not well constrained.  The goodness of fit depends largely upon the clustering of points and the error on the spectral index.

The above method was applied to the 17 events that showed a high magnitude of correlation coefficient in the electron velocity spectral index and the type III starting height.  There were 10 events which had physically meaningful predictions (positive values) for the starting height $h_{acc}$ and the vertical extent of the acceleration region $d$ (denoted by a full circle in Table \ref{tab:Events}). Again, the predictions were found from the intercept and gradient of the straight-line fit, respectively.  We have presented the data obtained for the flare acceleration region heights $h_{acc}$ and vertical extents $d$ in Table \ref{tab:acc_params}.  We find the mean $h_{acc}=98$~Mm and the mean $d=8.3$~Mm.  The spread of acceleration region values were quite high, with standard deviations of $53$~Mm and $3.3$~Mm respectively.  The acceleration region size is well defined from the fitting routine, with an error from the linear fit being at most  $16~\%$.  The acceleration region height is slightly less well defined for the flares with smaller estimated altitudes.  The largest uncertainty comes from flare on the 25th July 2004 when the errors are $40\%$ of the estimated altitude.  We must stress that the uncertainties shown are from the fit and give an indication as to the robustness of a linear fit to the data.  There are many more uncertainties in the estimation of the acceleration region parameters which arise from the number of assumptions that have been used (see Section \ref{discussion}).

Regarding the threshold used for the type III starting frequency, if we increase threshold values by 1 we change the values in Table \ref{tab:acc_params} for the acceleration region size by at most $10\%$.  For the acceleration region height the majority of the flares only change by at most $20\%$.  However, the flares on the 18th March 2003 and 25th July 2004 change by 42 and 45 percent respectively.  This is because of their smaller estimated altitude (discussed more fully in Section \ref{discussion}).  The correlation coefficient barely changes, by at most a $3\%$ decrease or a $6\%$ increase.  We cannot decrease the threshold values by one as noisy frequency channels stop our estimates for the starting frequencies tracking (by eye) the start of the type III bursts.

The other 7 events chosen from Table \ref{tab:Events} that were indicated by a hollow circle gave negative values for $h_{acc}$ from the straight line fitting to the data.  This corresponded to inferred altitudes below the solar surface that is clearly not physical.  The last two events (9th June 2003 and 22nd October 2003) shown in Figure \ref{fig:dsi} are such examples.  In six of the seven events where this occurred, some of the starting frequencies were $\geq 1$~GHz.  As we used the same density model for all of the events deduced from the observations of many type III bursts in the 450-150 MHz range \citep{Saint-Hilaire_etal2013} this density model may not be suited to type III bursts at high frequencies.  It would have been possible to adjust the density model by increasing the density scale height but we preferred to have the same density model in all of the events for a fair comparison.  We address the assumed density model further in Section \ref{discussion}.  The seventh event (29th September 2002) presented a non-robust correlation.  Increasing the threshold for the starting frequency from 2 to 3 decreased the correlation coefficient to 0.09.

\vspace{20pt}
\begin{center}
\begin{table}
\centering
\begin{tabular}{ c  c  c  c  c  c }

\hline\hline

 Date &  $h_{acc}$ [Mm] & $h_{acc}$ error [Mm] & $d$ [Mm] & $d$ error [Mm] \\ \hline
 
 20/02/02  &  $183$ &  $1$    &   $2.1$  &   $0.1$  \\
 20/04/02  &  $158$ &  $5$    &   $5.6$  &   $0.6$  \\
 02/06/02  &  $152$ &  $3$    &   $5.5$  &   $0.3$  \\
 19/07/02  &  $92$  &  $16$   &   $8.3$  &   $1$    \\
 10/09/02  &  $116$ &  $3$    &   $8.7$  &   $0.3$  \\
 14/09/02  &  $87$  &  $9$    &   $9.1$  &   $0.8$  \\
 10/03/03  &  $55$  &  $12$   &   $11$   &   $1.0$  \\
 18/03/03  &  $25$  &  $8$    &   $12$   &   $0.7$  \\
 12/06/03  &  $73$  &  $4$    &   $7.5$  &   $0.3$  \\
 25/07/04  &  $40$  &  $16$   &   $13$   &   $1$    \\
\hline
\end{tabular}
\vspace{20pt}
\caption{The derived values for the height and vertical extent of the acceleration region for ten events.  The one-sigma error is also given for each event for both values, from the linear fit to the data.}
\label{tab:acc_params}
\end{table}
\end{center}

\section{Discussion}\label{discussion}

The acceleration region characteristics found in Table \ref{tab:acc_params} are consistent to the ones deduced in \citet{Reid_etal2011}.  For the studied events, the results suggest an extended coronal acceleration region that is high in the corona.  One should keep in mind that we have analysed a subset of solar flares.  In our events, the flare accelerated electrons must propagate into the upper corona to produce emission between 450 and 150 MHz.  Confined flares which would be expected to have low lying acceleration regions are less likely to produce type III radio emission.  Moreover, all the flares were quite low energy in terms of their GOES class (mostly C class flares).  This was a direct result of selecting events where the only radio emission was type III bursts to avoid complicated radio dynamic spectra where the type III starting frequency would be ambiguous.

The height of the deduced acceleration regions are quite large with respect to the typical heights of soft X-ray loops \citep[e.g.][]{KontarJeffrey2010}.  This is partially a selection criteria based on the events that we have chosen since the analysed events are not confined flares and are associated with radio emissions starting around 600 MHz.  The other reason is that type III radio bursts originate quite high in the solar atmosphere.  The problem of having a low altitude acceleration region is the small distance an electron beam requires before any Langmuir wave instability occurs.

The rest of the discussion is broken down into three sections.  We first examine the robustness of our observed anti-correlation.  We then look at the assumptions that went into our model.  Finally we look at further extensions to our model that includes more physics.


\subsection{Uncorrelated events}

We now analyse the robustness of the anti-correlation that we found between the starting frequency of the radio bursts and the hard X-ray spectral indices.  Whilst we saw a good proportion (17/30) of events displaying this anti-correlation over the entire event, there were a number of events that had a correlation coefficient that was very small.  Why did we not observe an anit-correlation?  We suggest some of the reasons below.

\begin{figure*}
 \includegraphics[width=0.50\textwidth]{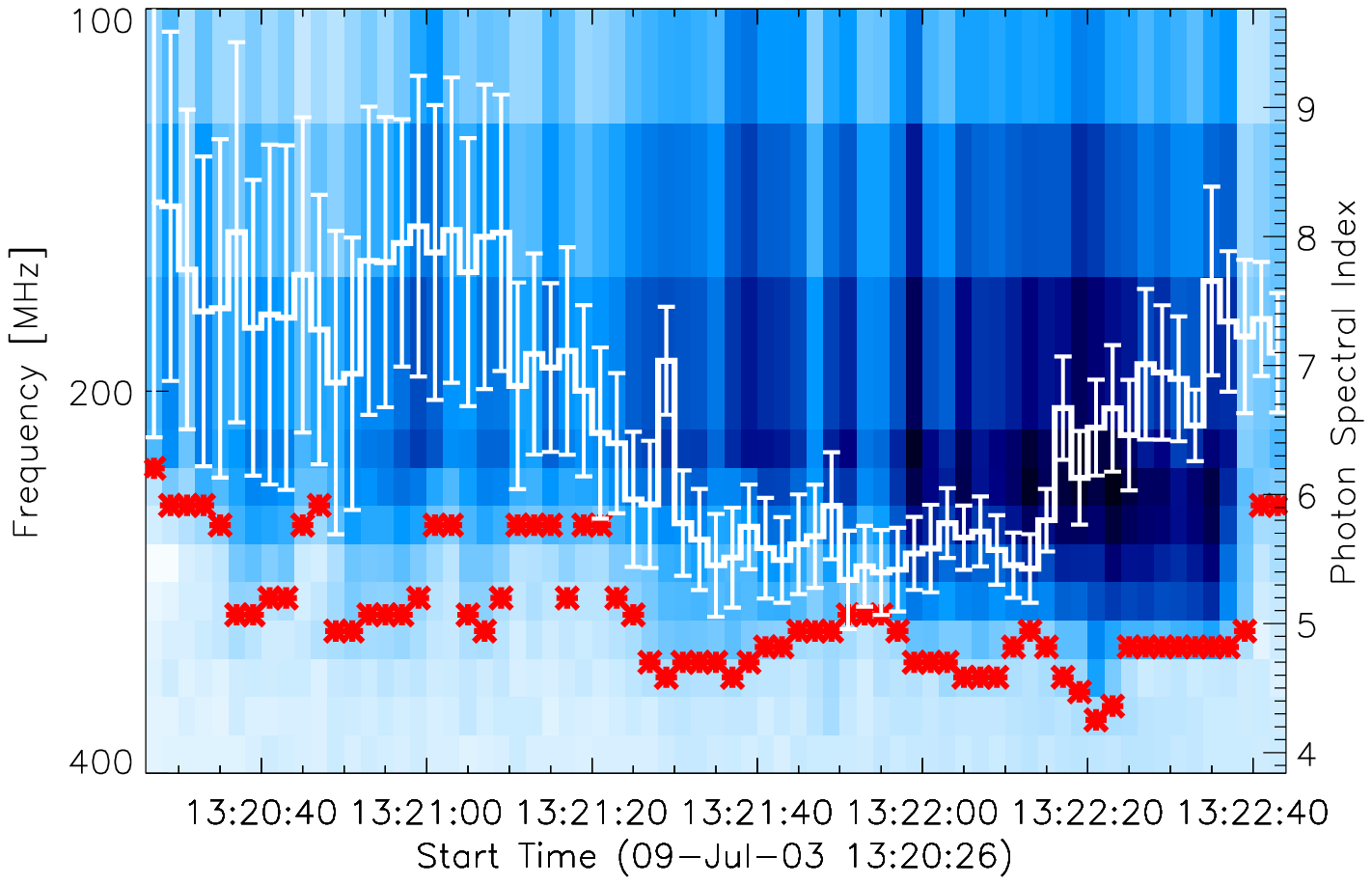}
 \includegraphics[width=0.50\textwidth]{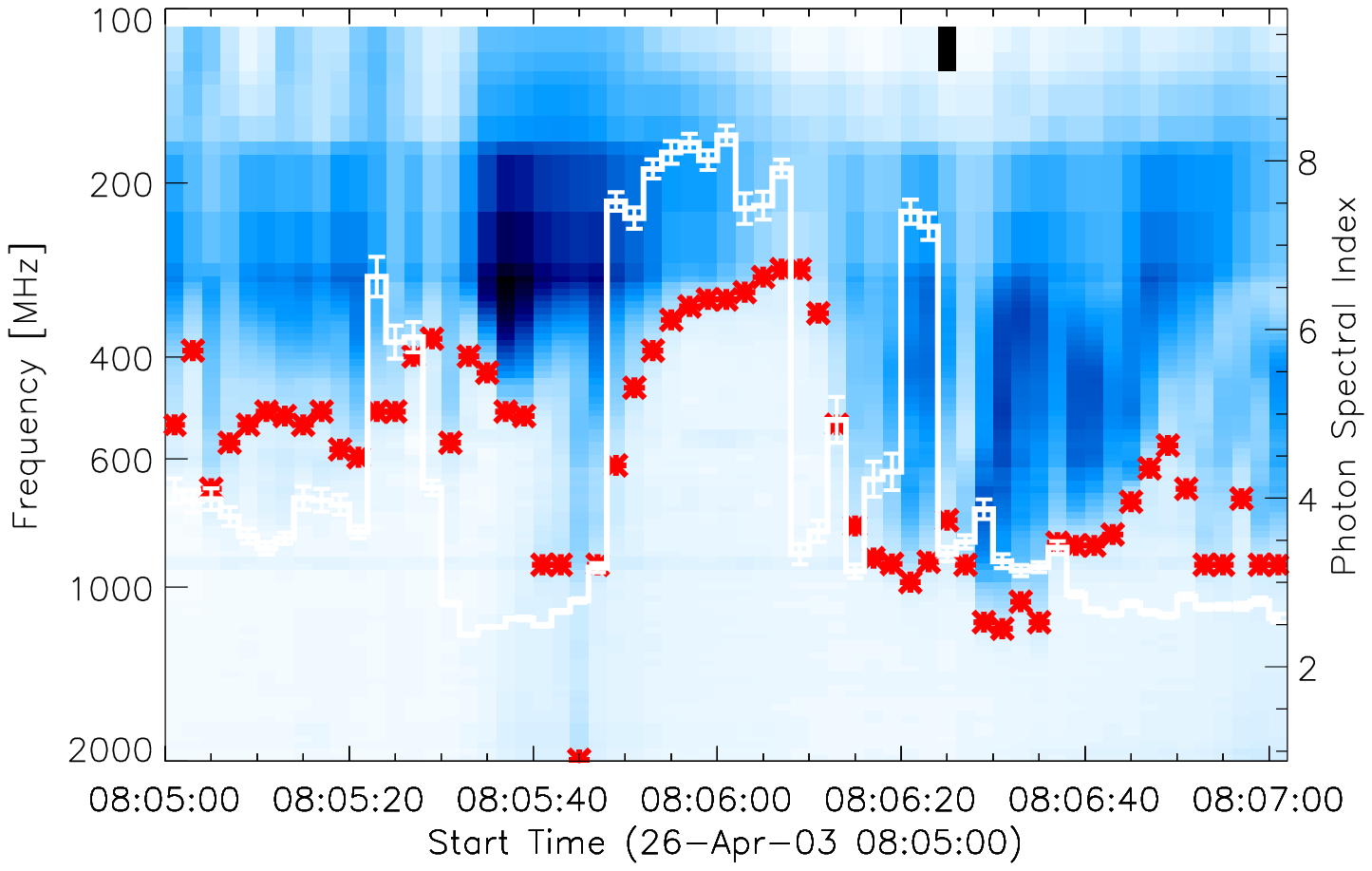}
 \includegraphics[width=0.50\textwidth]{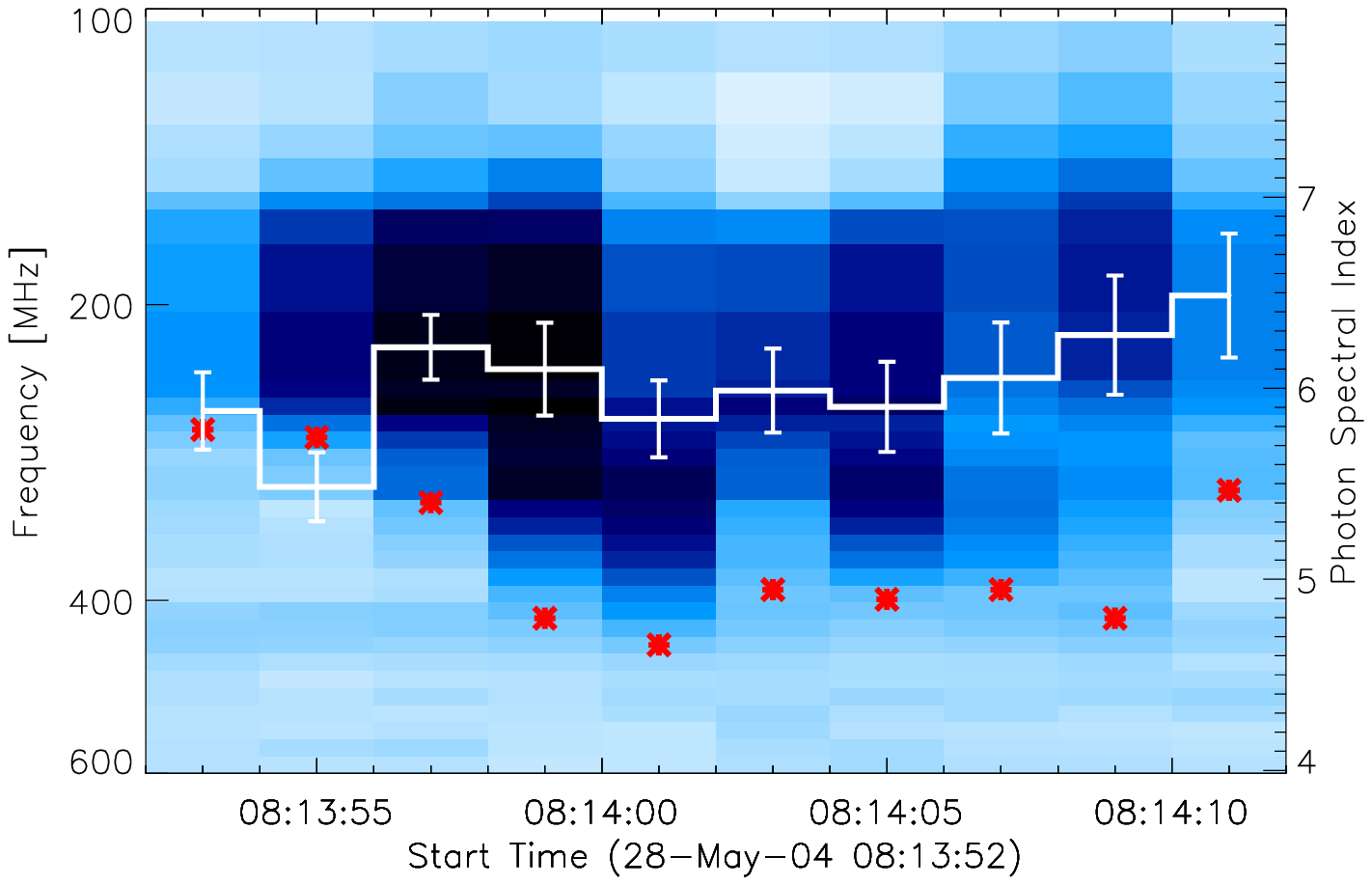}
 \includegraphics[width=0.50\textwidth]{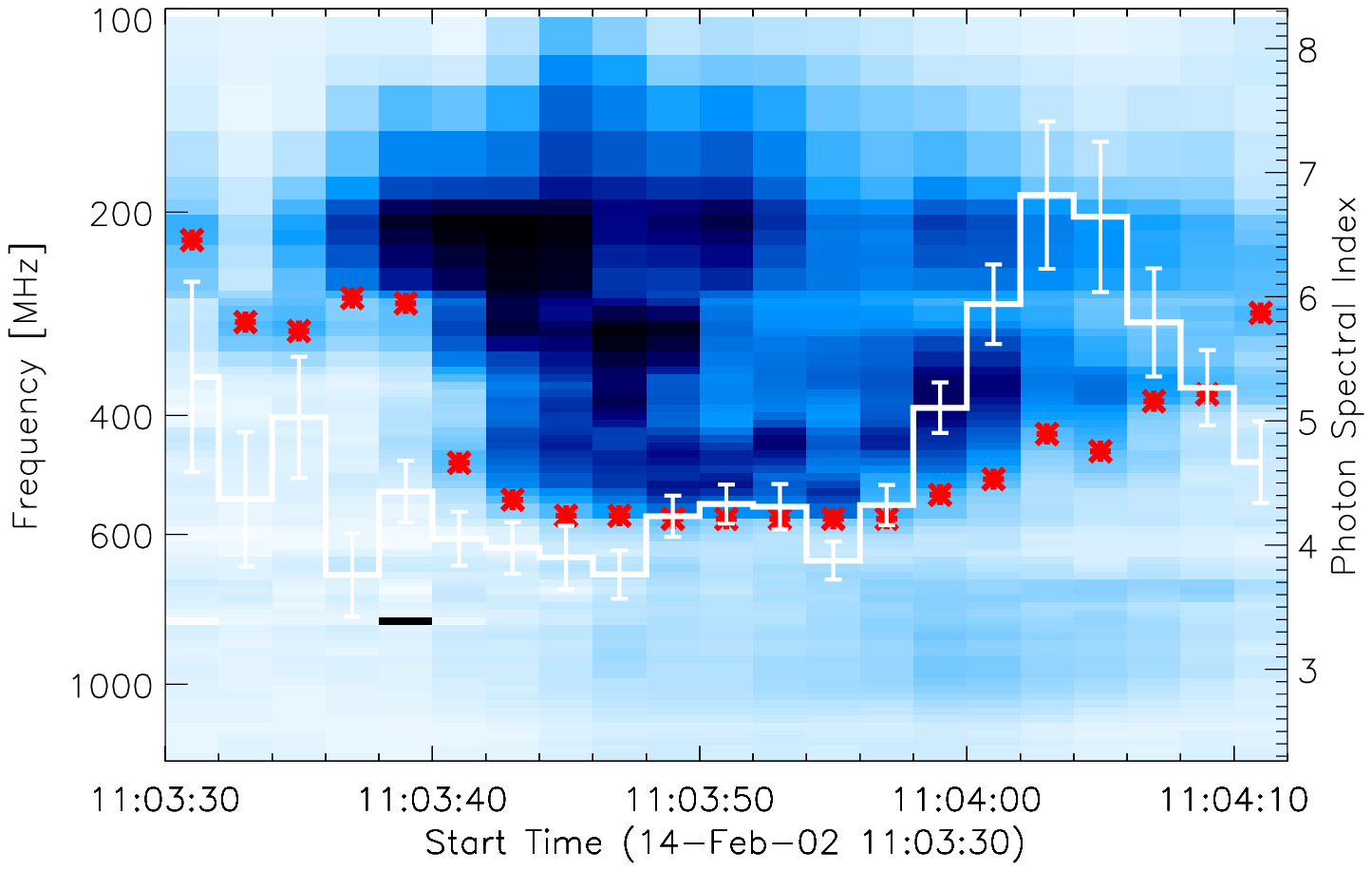}
\caption{Some example of events that did not show a relation between the groups of type IIIs starting frequencies and the X-ray spectral indices}
\label{fig:unrel}
\end{figure*}

There were a number of events with a very soft spectral index in X-rays and large error bars (9th July 2003 in Figure \ref{fig:unrel}).  We must question whether or not a non-thermal population of electrons generated the X-ray emission at high energies or whether the entire X-ray signature arises from a thermal population of electrons.  In the latter case we would not expect the X-ray signature to be related to the group of type IIIs.  

There were four events that had a longer duration (two minutes or greater) with respect to the other events.  These events showed some relation between the spectral index and the starting frequency, for example the event on the 26th April 2003 between 08:05:26 and 08:06:10 UT (Figure \ref{fig:unrel}).  However other times did not show any relation like the end of the event.  It could be that multiple acceleration regions were present for these flares or that the magnetic field structure was more complicated and/or changed dramatically during the event.   It means that for those events the link between HXR producing electrons and electron beams propagating up into the high corona is not a simple as the cartoon in Figure \ref{fig:overview}.  Let us recall that events in which the link between HXR emissions and radio decametric / metric emissions changes on timescales of a few minutes have been reported in the literature \citep[see e.g.][as an example]{Vilmer_etal2003}.

A number of the events were quite short in duration like the 28th May 2004 event in Figure \ref{fig:unrel}.  If there were one or two points that did not follow the trend (the first two points have a harder spectral index but not a higher starting frequency) then the correlation coefficient between the HXR spectral index and the starting frequency would be poor.  For these times electrons might not have been able to stream both up and down in coronal altitude leading to behaviour in radio that was not mirrored in X-rays (or vice-versa).

Finally there were events for which we did not see any connection between the X-ray signature and the group of type III bursts like the 14th February 2002 event in Figure \ref{fig:unrel}.  We had good signal to noise from the X-rays, and not too many anomalous points.  We see a decrease in spectral index at a similar time to an increase in the type III starting frequencies but the rise and decay of both wavelengths does not match.  In these situations the event may just not be as simple as we have assumed for our model.

\subsection{Analysis of assumptions} \label{assumptions}

We have used some assumptions to obtain the electron beam characteristics from the observed data.  One such assumption was the thick target model to obtain the electron velocity spectral index from the hard X-ray spectral index.  Between the electrons responsible for X-ray emission in the chromosphere and the electron beams detected in-situ near the Earth there exists a correlation in energy spectra \citep{Krucker_etal2007}.  The correlation does not coincide with the thick target model or the thin target model but lies somewhere in between.  Under this scenario we derive $\alpha =2\gamma$, so a scaler (2) would be subtracted from the velocity spectral indices we used in this study.  A correlation will still be expected between the hard X-ray spectral index and the type III starting frequency.  The effect would correspond to the height of the acceleration region changing by a very small amount (e.g. $h_{acc}$ increases by 16\% for the event on 19th July 2002). 

For a few events the spectrum would have been better approximated using a broken power-law with a break energy around 50 keV.  Again we mention the statistical correlation between the electron spectral index above the break energy deduced from X-ray observations and the measured spectral index above the break energy at 1~AU \citep{Krucker_etal2007}.  We also know from measurements of electron spectrum at 1~AU \citep{Krucker_etal2009} and modelling of electron transport from the Sun to the Earth \citep{KontarReid2009,ReidKontar2010,ReidKontar2013} that the spectral index below the break energy is linearly correlated with the spectral index above the break energy.  Therefore, a single power-law fit will not be changed substantially by a presence of a break in the spectrum.  Moreover, the increased complexity of using a broken power-law would introduce as much ambiguity as it solved.

We have assumed a fixed particle accelerator in space during the whole flare.  As mentioned in \citet{Reid_etal2011} , movement along the direction of travel at the Alfven speed is usually insufficient to explain the change in observed starting frequency of the type III bursts.  Even at 1 Mm s$^{-1}$, such motion could not account for the change in starting heights observed for the majority of events (see Figures \ref{fig:sfsi} and \ref{fig:dsi})  However, the particle accelerator does not need to be stationary in the direction perpendicular to the electron propagation.  The whole accelerator can jump/move from one field line to another.  Note that this is different from moving along the 1D field lines.  This will lead to different density profiles as a function of time assuming different `threads' of density profiles \citep{Kontar_etal2010}.  Motion of the accelerator in this way could be a reason we did not get a good correlation between the radio starting frequency and hard X-ray spectral index for some events.

\subsubsection{Flare morphology} \label{morphology}

\begin{figure*}
  \includegraphics[width=0.50\textwidth,trim=42 20 70 40,clip]{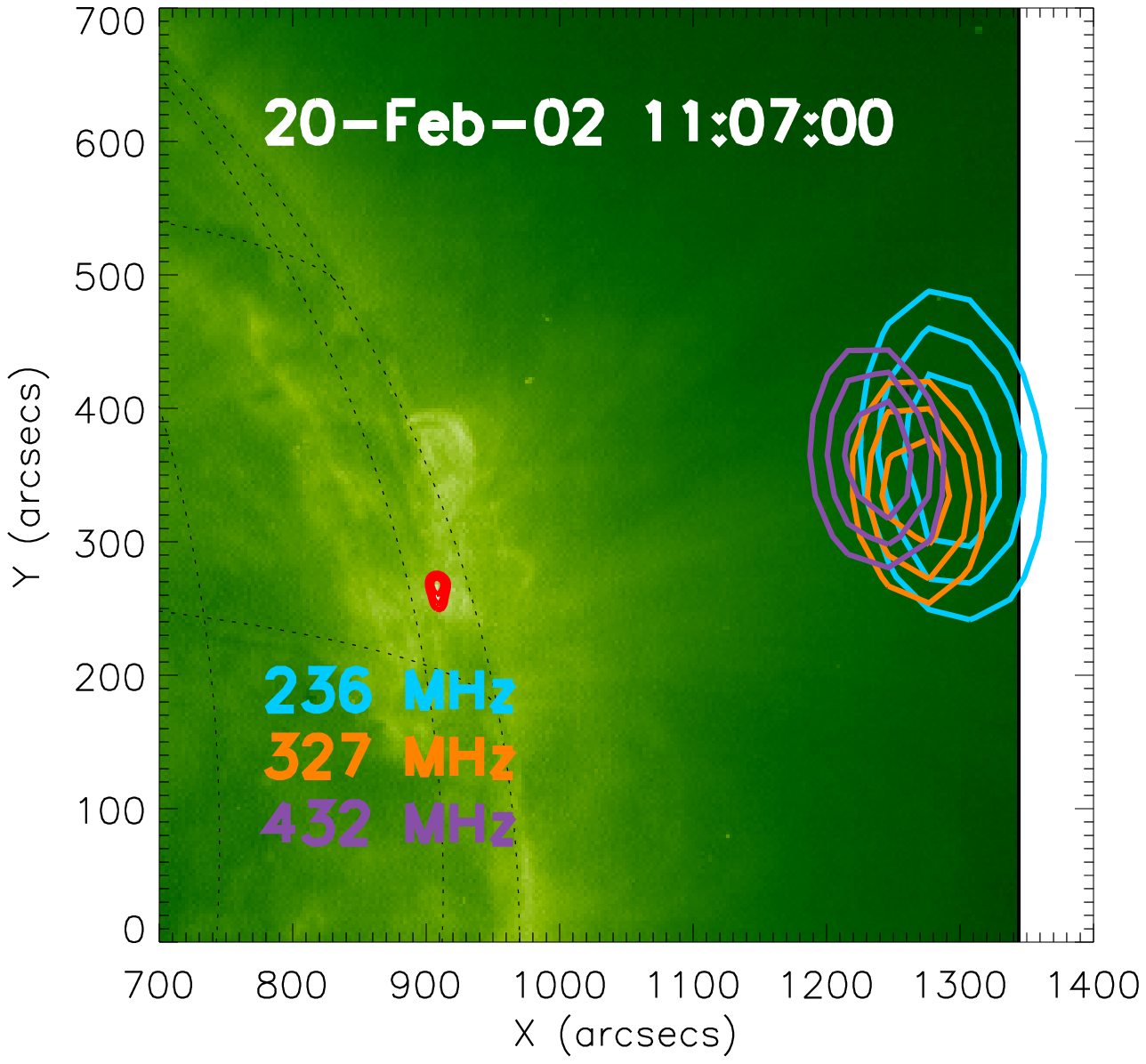}
  \includegraphics[width=0.50\textwidth,trim=42 20 70 40,clip]{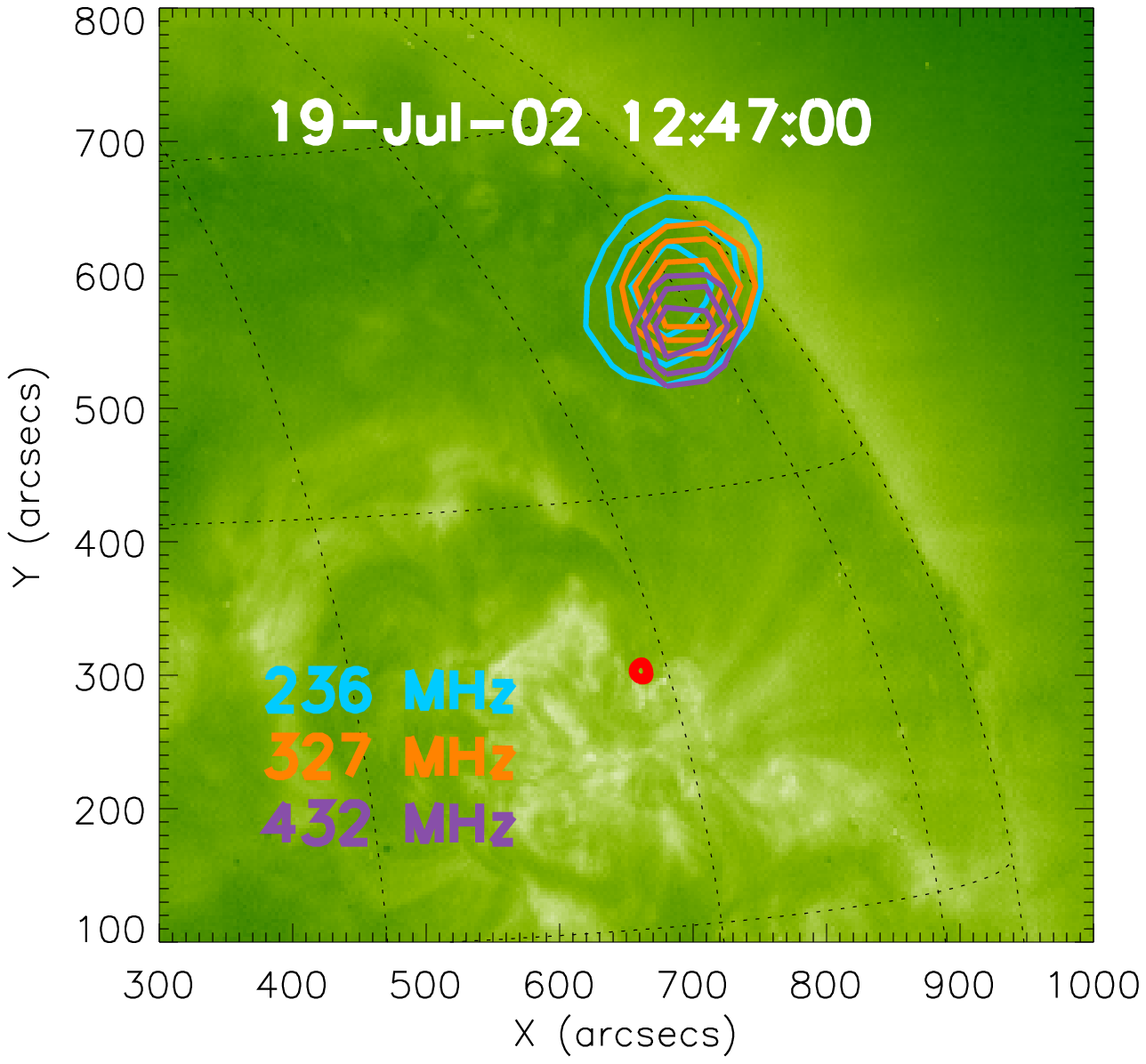}
  \includegraphics[width=0.50\textwidth,trim=42 20 70 40,clip]{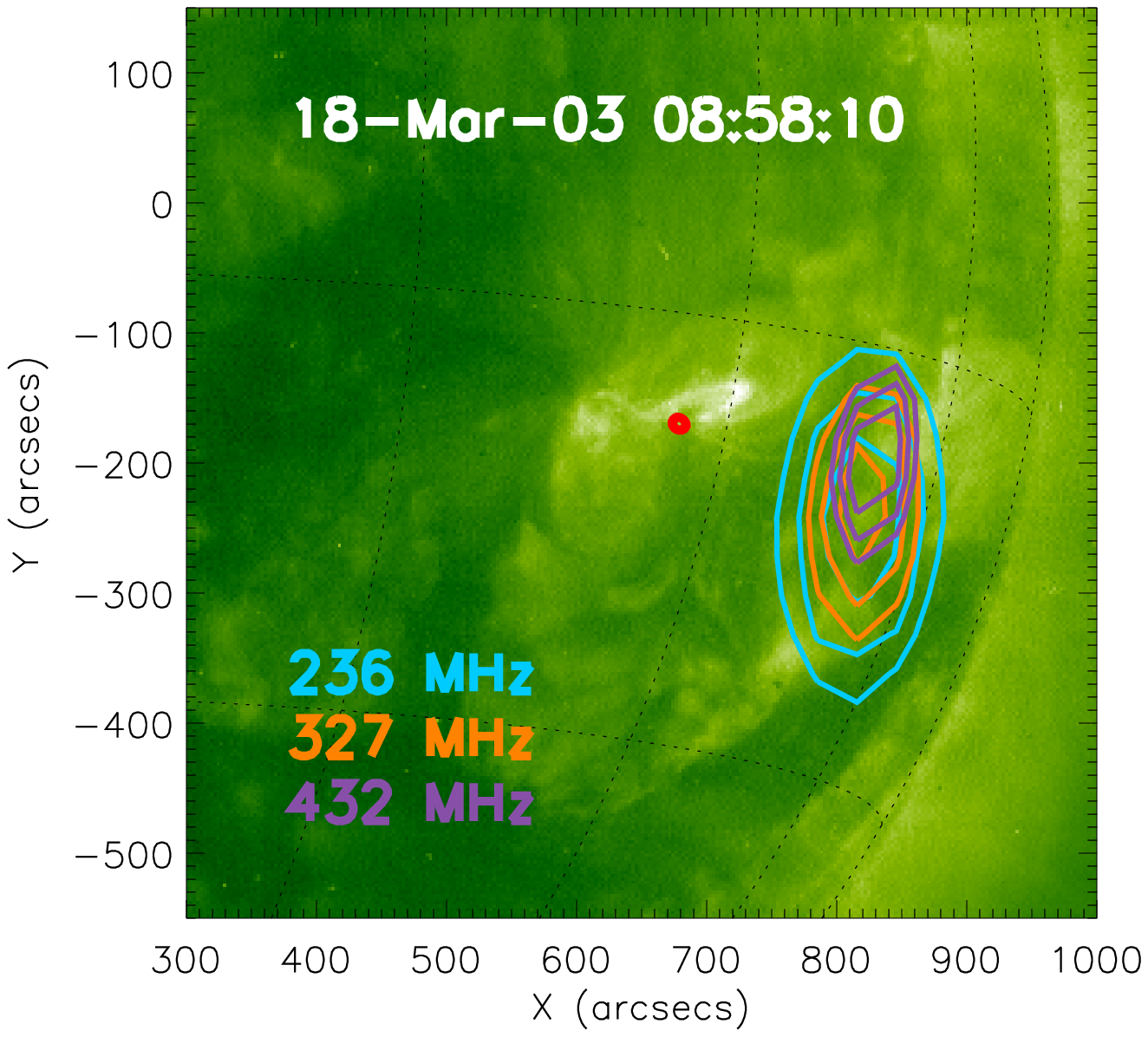}
  \includegraphics[width=0.50\textwidth,trim=42 20 70 40,clip]{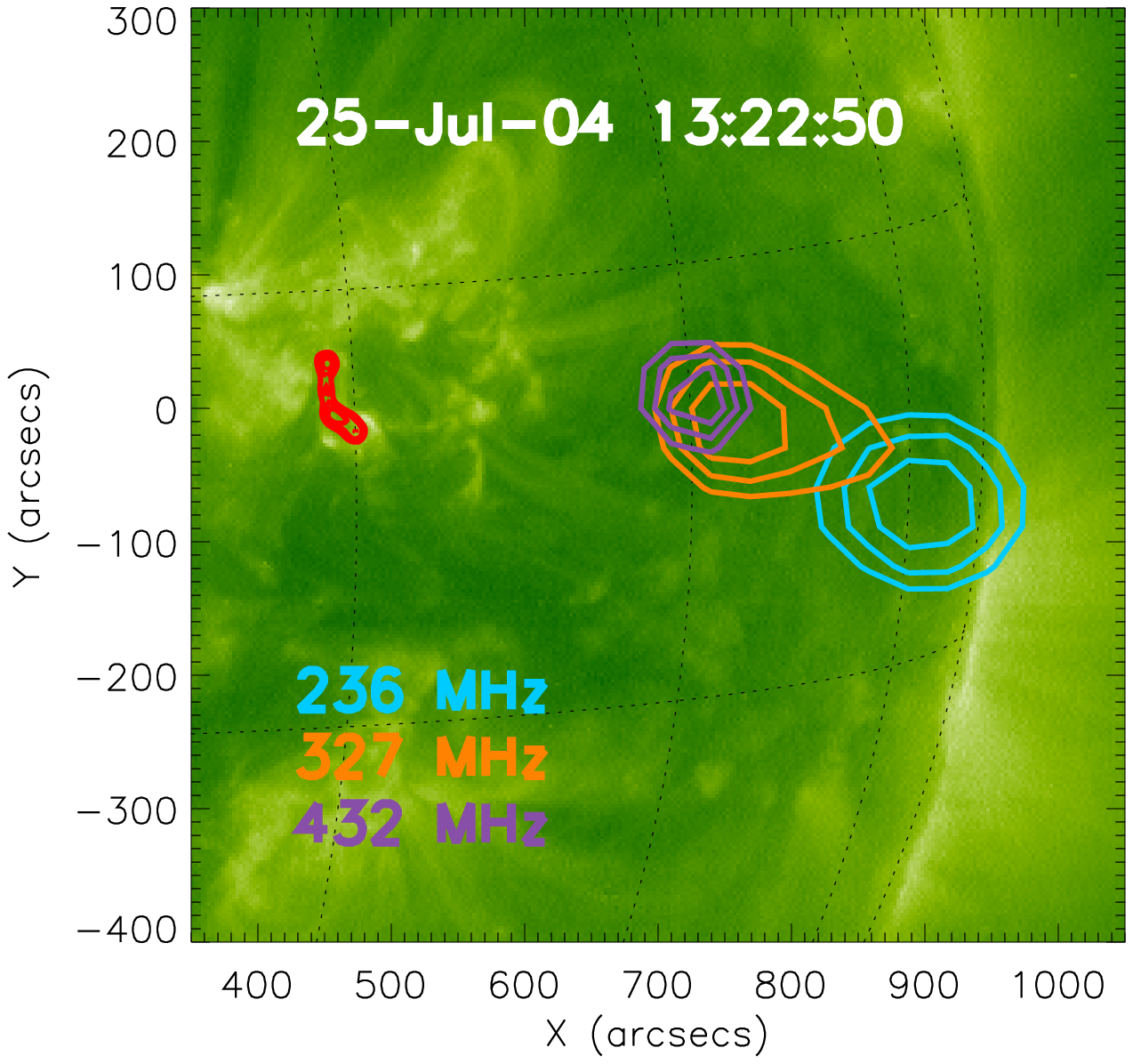}
\caption{Four flares that showed simple morphologies in the radio wavelength during the course of the type III bursts.  The flare is imaged in EUV $195~\text{\AA}$ by EIT.  At the base of the loops there are the RHESSI 12-25 keV X-rays (red) integrated over the entire event.  Higher in the corona there are the NRH 432~MHz (purple), 327~MHz (orange), and 237~MHz (blue) integrated over 10 seconds during the events where the start of this integration is indicated on each image.  All contours are at $10~\%$ intervals starting from 70 percent.}
\label{fig:morph}
\end{figure*}

As sketched in Figure \ref{fig:overview} and discussed in the introduction, the model used assumes a simple morphology for each of the flares.  We assume that the acceleration mechanism is producing bi-directional electron beams that emit X-rays in the low atmosphere and radio emission in the high atmosphere.  This assumption is in-line with the standard model for solar flares.  To check the morphology we looked at images of the 17 events that showed a good correlation between the starting heights and the spectral indices.  Again, we used the Nan\c{c}ay Radioheliograph to image the radio emission and RHESSI to image the X-ray emission.  We found that all 17 events showed radio sources near the X-ray sources.


For 11 of the 17 events a simple morphology was unambiguously observed.  For 7 of these 11 events with a simple morphology, parameters of the electron acceleration sites were derived.  The parameters of the acceleration regions derived for these 7 events show similar heights as the whole sample of events, with a mean of 89 Mm and a high standard deviation from this mean of 60 Mm.  For the vertical extent of the acceleration regions we again have similar values with a mean of 9 Mm and a standard deviation of 4 Mm.  

Four sample events are shown in Figure \ref{fig:morph}.  We can observe the high altitudes of the radio emission.  Given that the propagating electron beam can quickly become unstable from a small acceleration region, it becomes hard to explain high altitude radio emission from a low-lying acceleration region.

Of the remaining 6 events, 4 showed two or more sources of radio emission at different positions in the same frequency for the same point in time.  It is very possible that some flares have more than one avenue of escape for the electron beams and hence we can observe radio emission originating from two different points in the atmosphere.  The single acceleration region may still be viable for these events but the complicated nature of their morphology puts into question some of the assumptions the work is based upon.  The last 2 events did not have any Nan\c{c}ay Radioheliograph data.



We should note that the radio (and to a lesser extent the hard X-ray) sources for all events develop over time.  Their positions change during the course of the flare \citep[see][as an example for the 20th February 2002 flare]{Vilmer_etal2002}.  However, the events that show the simple scenario for the entire duration keep the simple structure of having one strong source in each frequency band and have an observable progression from high to low frequencies.





\subsubsection{Density model}

\begin{figure}
  \includegraphics[width=0.96\columnwidth]{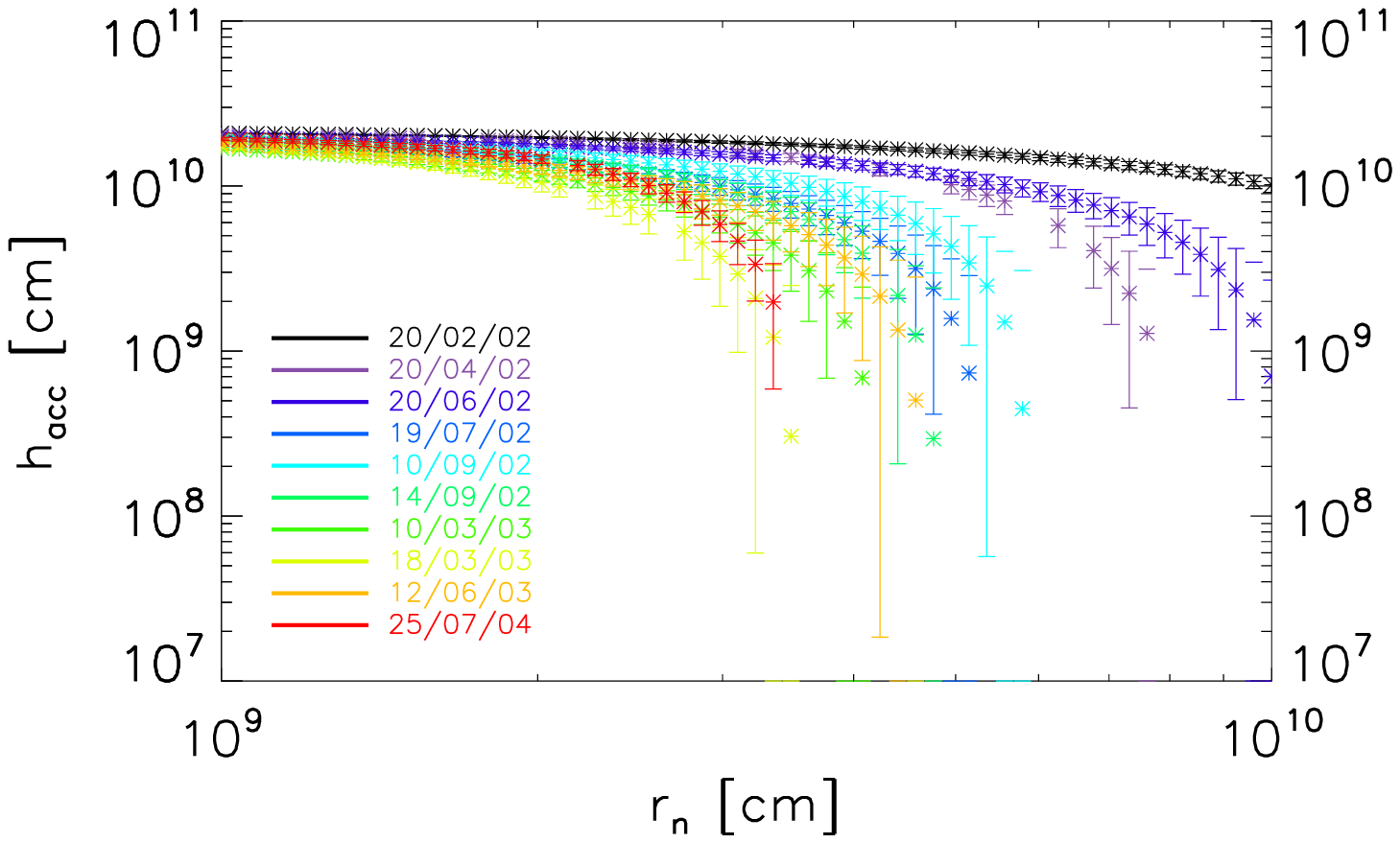}
  \includegraphics[width=0.96\columnwidth]{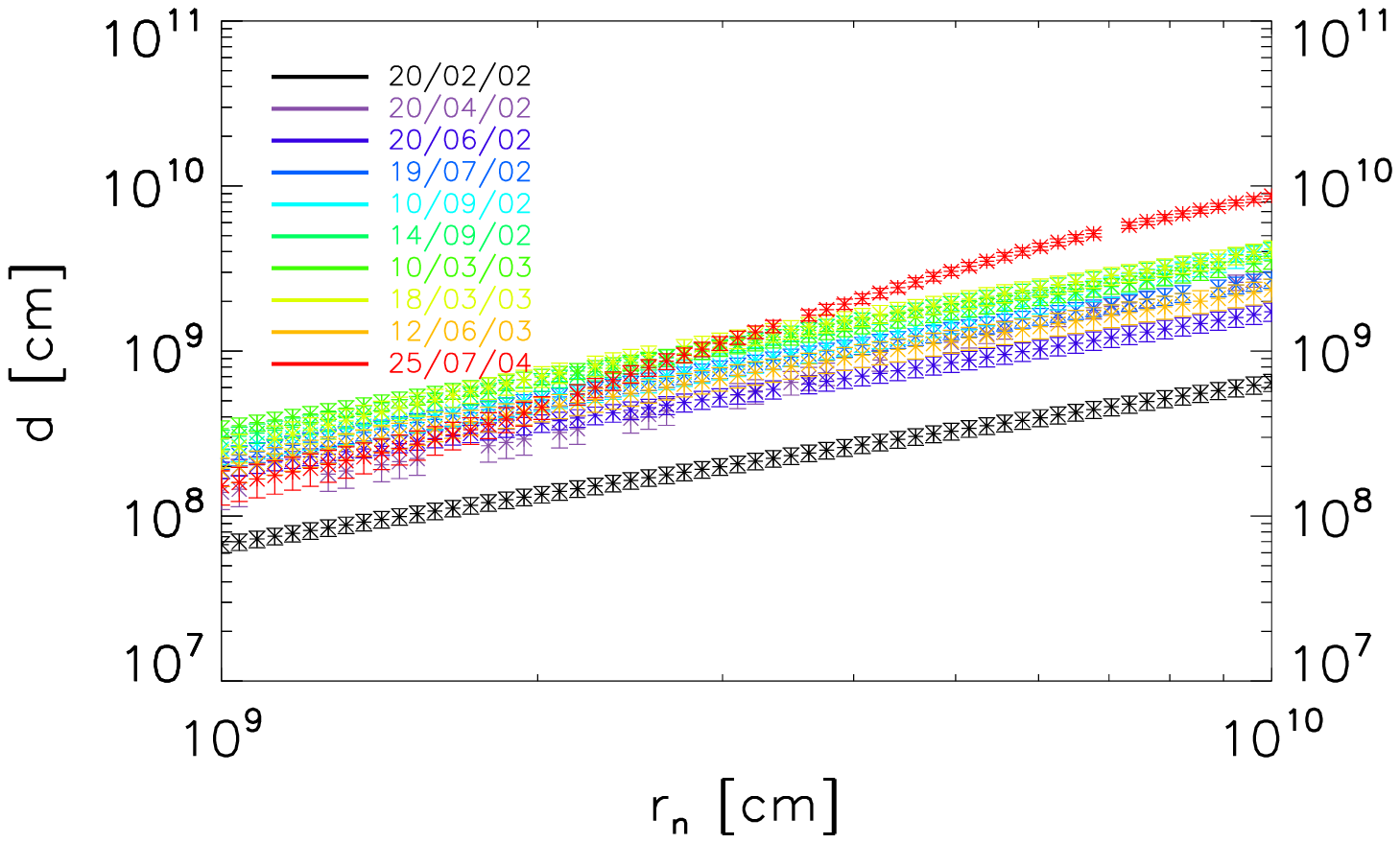}
\caption{The estimated acceleration region parameters of $h_{acc}$ (top) and $d$ (bottom) for different density models.  The characteristic scale of the density model, $r_n$ has been varied as $10^9 \leq r_n \leq 10^{10}$~cm.  The ten events that had good predictions of acceleration region characteristics are shown in different colours.}
\label{fig:dens_comp}
\end{figure}

Another important assumption we used was the exponential density model for the corona.  We can change the form of the exponential density model in two ways.  First, we can change the reference values $h_0$ and $n_0$.  Doing so modifies $A$ in Equation (\ref{exp_dens}).  If we modify $A$, the height of the estimated acceleration region will change linearly but the vertical extent of the acceleration region will remain the same.  

We can also change the density model by modifying the gradient $dn/dr$ of the density model.  To do this we modify $r_n$, the characteristic scale of the density model.  Figure \ref{fig:dens_comp} shows the change to $h_{acc}$ and $d$ if we modify $r_n$ (modifying the density gradient).  For $h_{acc}$ the limit of all events as $r_n \rightarrow 0$ is the reference height $h_0=2.21\times10^{10}$~cm (Equation (\ref{exp_dens})), obtained from \citet{Paesold_etal2001}.  This occurs because as $r_n \rightarrow 0$, $dn/dr \rightarrow \infty$.  Zero variation of height with density means that all acceleration regions would be situated at the same height BUT we would not see any type III bursts!  When $r_n$ is increased we get a turnover that corresponds to the acceleration region being situated closer to the solar surface.  If $r_n$ is too high then our model predicts $h_{acc}$ is less than $d$, essentially giving undefined acceleration region characteristics.  For the size of the acceleration region $d$, any change in $r_n$ creates a proportional change in $d$.  However, we cannot vary the characteristic scale of the density model much if we want to keep a realistic coronal density gradient.  Any corresponding change in the values of $h_{acc}$ and $d$ will therefore be small.

Changing our density model by decreasing $r_n$ to $10^9$~cm we model a steeper spatial gradient in the electron density.  This is perhaps a more realistic model for the events where we observe starting frequencies in the GHz.  Indeed we are able to obtain sensible acceleration region heights for three of the six events that we were unable to using $r_n=10^{9.5}$~cm derived from the NRH observations \citep{Saint-Hilaire_etal2013}.  The other events may have had either anomalous data points like the lowest starting height on the 22nd October 2013 (see Figure \ref{fig:dsi}) or simply an exponential density model was not consistent with the coronal conditions at the time.

\subsection{Analysis of the Model}

The model we have assumed is relatively simple.  That we find a correlation between the starting height of the type III bursts and the electron beam spectral indices derived from the X-ray observations using only a simple model gives us confidence that these two variable are connected.  We now examine how adding additional physics to the model will change the estimations of the electron beam characteristics.


\subsubsection{Gaussian injection}

As \citet{FarnikKarlicky2007} mention with respect to X-ray flares and reverse type III bursts, the initial distribution function used for the electron beam is important.  We can change our initial distribution from an exponential function to a Gaussian of the form 
\begin{equation}
f(r,v,t=0) \propto \exp(-r^2/d^2).
\end{equation} 
A change in the initial distribution function will change at what distance we observe an unstable electron beam.   The differential in velocity space of the distribution function changes and hence we end up changing Equation (\ref{eq_mx_c}) to \citep{MelnikKontar1999,Reid_etal2011} 
\begin{equation}\label{eq_mx_c_gauss}
h_{typeIII}=d\sqrt{\alpha/2}+h_{acc}.
\end{equation}
We note a misprint in \citet{Reid_etal2011} where the equation for $h_{typeIII}$ was incorrectly given as $h_{typeIII}=2d\sqrt{\alpha}+h_{acc}$.  For this new fit some of the flares give undefined results for the acceleration region properties (e.g. acceleration region sizes larger than the heights).  Others show that the acceleration region height decreases while the vertical extent of the acceleration region increases.  This trend can be explained because the spectral index becomes less and less influential on the evolution in the starting height the more peaked an initial distribution becomes.  For example in Equation (\ref{eq_mx_c_gauss}) only the square root of the spectral index is important in determining the starting frequency.  To explain the observed variation the size of the acceleration region must become larger and hence the height of the acceleration region becomes lower.  We also note that the correlation between the starting height of the type III bursts and the electron velocity spectral index is changed very slightly when we consider Equation (\ref{eq_mx_c_gauss}) but only by a maximum of 0.01.

\subsubsection{Injection with pitch angle}

We can also change the initial distribution function by considering the pitch angle distribution of the electrons.  Rather than initially assuming a beam travelling solely along the direction of propagation (1D) we instead assume a distribution function of the form
\begin{equation}
f(v,r,\mu,t=0) = g_0(v)\exp(-|r|/d)\psi(\mu)
\end{equation}
where $\psi(\mu)$ is the pitch angle distribution and $\int_{-1}^1 \psi(\mu)d\mu = 1$.  We consider an initial isotropic distribution function where $\psi(\mu)=\rm{const}$ at $t=0$.  By integrating over pitch angle to get the reduced distribution (see Appendix \ref{pitch}) we find that Equation \ref{eq_mx_c} changes to
\begin{equation}\label{eq_mx_c_mu}
h_{typeIII}=d(\alpha+1)+h_{acc}.  
\end{equation}
We observe that Equation (\ref{eq_mx_c_mu}) is only different to Equation (\ref{eq_mx_c}) by an addition of one to the spectral index $\alpha$.  As discussed in Section \ref{assumptions} regarding the thick target model, a small, static difference to $\alpha$ will only affect the starting height of the acceleration region.  It will not affect the size of the acceleration region nor will it affect the correlation coefficient.

\subsubsection{Time injection}

The third way we can change the electron injection is to have it vary as a function of time.  We have previously been considering an injection that is a delta function in time (instantaneous).  We now consider a source function for the electron injection of the form 
\begin{equation}
S(v,r,t) = g_0(v)\exp(-|r|/d)\exp(-|t-t_0|/\tau)
\end{equation}
where $\tau$ is the characteristic timescale of the injection and the peak of the injection occurs when $t=t_0$.  This is a similar form to what was assumed in \citet{ReidKontar2013}.  We can approximate the instability distance by using  $\Delta r \approx (d+v\tau)\alpha$.  For beam velocities of $10^{10}~\rm{cm~s}^{-1}$ and acceleration scales of $10^9~\rm{cm}$ we find that acceleration characteristic times over 0.1 second will start to become important and lengthen the instability distance with respect to an instantaneous injection.  However, we observe 0.1 s time structures in both X-ray \citep{Kiplinger_etal1983,Dennis1985,Vilmer_etal1994,Dmitriev_etal2006} and radio \citep[e.g.][]{CorreiaKaufmann1987,Aschwanden_etal1995b,Kaufmann_etal2009} observations.  Multiple accelerated electron beams that overlap in space along the same field lines would also cause a fast injection profile to affect the type III starting frequency.  A slow injection profile on the order of a second or more could result in much smaller acceleration regions than we deduce in this study.























\subsubsection{Density fluctuations}

The level of Langmuir waves can be altered not just by the form of the accelerated electron beam but also by conditions in the ambient background plasma.  Fluctuations in the electron density are known to reduce the total level of Langmuir waves that are induced by an electron beam \citep[e.g.][]{Ryutov1969,SmithSime1979,Kontar2001,Li_etal2006,ReidKontar2013,Ziebell_etal2011,Li_etal2012} and matter in both the parallel and perpendicular direction of electron propagation \citep{Ratcliffe_etal2012}.  Wave refraction causes the Langmuir waves to be shifted out of resonance with the electron beam, suppressing the bump-in-tail instability.  The presence of substantial density fluctuations in the coronal plasma would reduce the growth rate of Langmuir waves.  This could alter the height and/or size of our predicted acceleration region characteristics.  The exact level of density fluctuations is found between $5-15\%$ in the slow solar wind \citep[e.g.][]{Woo_etal1995,Spangler2002,SpanglerSpitler2004} with different power-law density spectra at different scales \citep[e.g.][]{Celnikier_etal1987,Chen_etal2013} although \citet{Woo_etal1995} found a much reduced level of fluctuations very close to the Sun in the fast solar wind.

\section{Conclusion} \label{conclusion}

We have looked at a series of type III bursts and hard X-ray flares from a period of 6 years.  When the two emissions occur at the same time, the X-ray flux above 25 keV is high enough for detection, and the event duration is 20 seconds or longer we found a correlation between the starting height of the groups of type III radio bursts and the spectral index of the energetic electrons in approximately 50$\%$ of events.  Moreover, some events that did not show a significant correlation for the entire flare duration showed trends in the starting frequency of the groups of type IIIs that was mirrored in the X-ray spectral indices over shorter times.  This correlation is expected to be harder to detect with higher energy flares on account of more complicated radio signatures in frequency space.

We observed a tendency for the starting frequency of the type III bursts to display a low-high-low trend.  Starting at a low frequency, the starting frequency became larger during the impulsive peak of the flare and then went lower during the decay of the flare.  This is analogous to the soft-hard-soft nature of the X-ray spectral index.

We used a model for the electron transport to predict the heights and vertical extents of the flare acceleration regions.  With a couple of assumptions we deduce parameters of the electron acceleration sites from the observable radio and X-ray emissions.  Deduced altitudes ranged from 183 Mm to 25 Mm with a mean of 100 Mm.  The vertical extent of the acceleration regions ranged from 16 Mm to 2 Mm with a mean vertical extent of 8 Mm.  The result was found for flares with a GOES class of M, C or B that caused only type III radio emission at decimetric wavelengths.  We finally note that the acceleration heights are as a mean larger than the ones deduced from X-ray analysis alone \citep{Aschwanden_etal1998}.  It must be noticed, however, that the flares analysed here are not confined flares which potentially explains the higher altitude heights. 



Our study is effective at inferring acceleration region characteristics that are largely unavailable but very important to our understanding of flare physics.  What is particularly key for future work is the inclusion of imaging at many more radio frequencies than are currently available.  With imaging in a wide frequency range we would be able to detect, in some events, the upward and downward propagating electron beams responsible for type IIIs and reverse type IIIs.  Detection of lower intensity X-rays would also shed more light on the spatial characteristics of flares acceleration regions as we could obtain the electron beam spectral index from the weak coronal emission produced by upward travelling electron beams.

\begin{acknowledgements}
The European Commission is acknowledged for funding from the HESPE Network (FP7-SPACE-2010-263086).  Hamish Reid acknowledges the financial support from a SUPA Advanced Fellowship.  Nicole Vilmer acknowledges support from the Centre National d'Etudes Spatiales (CNES) and from the French program on Solar-Terrestrial Physics (PNST) of INSU/CNRS for the participation to the RHESSI project.  Eduard Kontar acknowledges financial support from a STFC rolling grant. Financial support by the Royal Society grant (RG090411) is gratefully acknowledged. The overall effort has greatly benefited from support by a grant from the International Space Science Institute (ISSI) in Bern, Switzerland.  Collaborative work was supported by a British council Franco-British alliance grant.  The NRH is funded by the French Ministry of Education and the R\'{e}gion Centre.  The Institute of Astronomy, ETH Zurich and FHNW Windisch, Switzerland is acknowledged for funding Phoenix 2.
\end{acknowledgements}

\bibliographystyle{aa}
\bibliography{xradio2}

\appendix

\section{Electron beam instability} \label{app1}

We initially assume a one dimensional electron beam injected at height $r=0$ of the form \citep[see][]{Reid_etal2011}
\begin{eqnarray}\label{eq_init_f}
f(v,r,t=0)= g_0(v)exp(-|r|/d),
\end{eqnarray}
where $g_0(v) \propto v^{-\alpha}$ and $\alpha$ is the spectral index of the electron beam (in velocity space).  $d$ is the characteristic size of the electron beam (and consequently the size of the acceleration region).  At $t>0$ the electron beam propagates through space (reaching distance $vt_1$ at time $t_1$), creating a bump-in-tail distribution that causes resonant Langmuir wave growth as $\partial f/\partial v > 0$ \citep{DrummondPines1962,Vedenov_etal1962}.  The Langmuir wave quasilinear growth rate $\gamma(v,r)$ and the collisional absorption of Langmuir waves $\gamma_c$  are given by
\begin{eqnarray}\label{growth_rate}
\gamma(v,r) = \frac{\pi \omega_{pe}(r)}{n_e(r)}v^2\frac{\partial f}{\partial v}>\gamma_c,  \quad\quad\quad \gamma_c=\frac{\pi e^4 n_e(r)}{m_e^2v_{Te}^3}\ln \Lambda
\end{eqnarray}
where $\ln \Lambda $ is the Coulomb logarithm, taken as 20 for the parameters in the corona.  $\omega_{pe}(r)$, $n_e(r)$ and $v_{Te}$ are the background plasma frequency, density and thermal velocity respectively.  When the growth of Langmuir waves becomes larger than the collisional absorption of Langmuir waves from the background plasma we can obtain Langmuir waves orders of magnitude above thermal levels.  Radio waves can then be produced through wave-wave interactions at the local plasma frequency and the harmonics, which we observe as type III radio bursts \citep[e.g.][]{KontarPecseli2002,Li_etal2008}.

At time $t>0$ we can describe the distribution function assuming no energy loss using
\begin{eqnarray}
f(v,r,t)=g_0(v)exp(-|r-vt|/d)
\end{eqnarray}
and the growth rate for Langmuir waves becomes
\begin{eqnarray}
\gamma (v)=\frac{\pi \omega_{pe}(r)}{n_e(r)}v^2
f(v,r,t)\left(\frac{t}{d}-\frac{\alpha}{v}\right).
\label{eq_gamma}
\end{eqnarray}
Langmuir waves will occur at a height $h_{typeIII}$ which can be found at a distance $\Delta r = h_{typeIII} - h_{acc}$ from the acceleration site at $h_{acc}$.  Using the condition that Langmuir waves will be prolific when the growth rate exceeds the collisional absorption rate we can find the distance $\Delta r$ via
\begin{eqnarray}
\Delta r=d\left(\alpha + \frac{\gamma_c n_e(r)}{\pi \omega_{pe}(r)}(v g_0(v))^{-1}\right).
\label{eq_x}
\end{eqnarray}
The second term in the brackets can be approximated using coronal parameters.  We can use $v g_0(v)=n_b$ were $n_b$ is the electron beam density.  We assume $n_e(r)=10^9~\rm{cm}^{-3}$, $T_e=2~\rm{MK}$, $n_b=10^4~\rm{cm}$ and we find that this term is around $10^{-3} \ll \alpha$.  Thus we find the simple relation
\begin{eqnarray}\label{eq_mx_c_appendix}
h_{typeIII}=d\alpha + h_{acc},
\end{eqnarray}
that equates the known quantities $h_{typeIII}$ and $\alpha$ we can deduce from observations to the unknown quantities of $d$, the vertical extent of the acceleration region and $h_{acc}$ the height of the acceleration region.

\section{Isotropic electron beam instability} \label{pitch}

We now assume an initial electron beam that can vary with pitch angle such that at $t=0$ we inject
\begin{equation}
f(v,r,t=0,\mu) = g_0(v)\exp(-|r|/d)\psi(\mu)
\end{equation}
where $\psi(\mu)$ is the pitch angle distribution and $\int_{-1}^1 \psi(\mu)d\mu = 1$.  We consider an initial isotropic distribution function where $\psi(\mu)=\rm{const}$ at $t=0$.  Again $g_0(v) \propto v^{-\alpha}$ and $\alpha$ is then spectral index of the electron beam (in velocity space).  $d$ is the characteristic size of the electron beam (and consequently the size of the acceleration region).  The distribution function at $t>0$ becomes
\begin{equation}
f(v,r,\mu,t) = g_0(v)\exp(-|r-v\mu t|/d)
\end{equation}
where $v$ is the speed of the electrons.  We can find the reduced distribution along $\vec{B}$ using 
\begin{align}\label{reduced_mu}
\begin{split}
f_{\parallel}(v,r,t)    & = \int_0^1{\mu f(v,r,\mu,t)d\mu} \\
         & =  g_0(v)\left[ \left(\frac{d}{vt}\right)-\left(\frac{d}{vt}\right)^2\right]\exp\left(\frac{-|r-v t|}{d}\right) \\
            & \quad + g_0(v)\left(\frac{d}{vt}\right)\exp\left(\frac{-|r|}{d}\right),
\end{split}
\end{align}
where we integrated for $r-v\mu t > 0$ as we are interested in the growing part of the electron beam where $\partial f/ \partial v> 0$.  Assuming that $r\gg d$ and $d/vt \ll 1$, as predicted by Equation (\ref{eq_mx_c}), we can approximate Equation (\ref{reduced_mu}) as 
\begin{align}\label{reduced_mu2}
\begin{split}
f_{\parallel}(v,r,t)    & \cong g_0(v)\left(\frac{d}{vt}\right)\exp\left(\frac{-|r-v t|}{d}\right) \\
                  & \cong Cv^{-(\alpha+1)}\left(\frac{d}{t}\right)\exp\left(\frac{-|r-v t|}{d}\right).
\end{split}
\end{align}
By finding $\partial f_{\parallel} / \partial v$ we obtain the equation for the growth rate of Langmuir waves
\begin{eqnarray}
\gamma (v)=\frac{\pi \omega_{pe}(r)}{n_e(r)}v^2
f_{\parallel}(v,r,t)\left(\frac{t}{d}-\frac{(\alpha+1)}{v}\right).
\end{eqnarray}
Using the same analysis demonstrated between Equations (\ref{eq_gamma}) and (\ref{eq_mx_c_appendix}) we find the simple relation
\begin{eqnarray}
h_{typeIII}=d(\alpha+1) + h_{acc},
\end{eqnarray}
that equates the known quantities $h_{typeIII}$ and $\alpha$ we can deduce from observations to the unknown quantities of $d$, the vertical extent of the acceleration region and $h_{acc}$ the height of the acceleration region.

\end{document}